# Jupiter's South Equatorial Belt cycle in 2009-2011: II. The SEB Revival

**John H. Rogers**          *A report of the Jupiter Section, using data from the JUPOS team*

______________________________________________________________________

## Abstract

A Revival of the South Equatorial Belt (SEB) is an organised disturbance on a grand scale.  It starts with a single vigorous outbreak from which energetic storms and disturbances spread around the planet in the different zonal currents.  The Revival that began in 2010 was better observed than any before it.  The observations largely validate the historical descriptions of these events: the major features portrayed therein, albeit at lower resolution, are indeed the large structural features described here.  Our major conclusions about the 2010 SEB Revival are as follows, and we show that most of them may be typical of SEB Revivals.
1)  The Revival started with a bright white plume.
2)  The initial plume erupted in a pre-existing cyclonic oval ('barge').  Subsequent white plumes continued to appear on the track of this barge, which was the location of the sub-surface source of the whole Revival.
3)  These plumes were extremely bright in the methane absorption band, i.e. thrusting up to very high altitudes, especially when new.
4)  Brilliant, methane-bright plumes also appeared along the leading edge of the  central branch.  Altogether, 7 plumes appeared at the source and at least 6 along the leading edge.
5)  The central branch of the outbreak was composed of large convective cells, each initiated by a bright plume, which only occupied a part of each cell, while a very dark streak defined its west edge.
6)  The southern branch began with darkening and sudden acceleration of pre-existing faint spots in a slowly retrograding wave-train.
7)  Subsequent darker spots in the southern branch were complex structures, not coherent vortices.
8)  Dark spots in the southern branch had typical SEBs jetstream speeds but were unusually far south.  This suggests either a complex vertical structure of the SEBs jet, or a real acceleration westwards on the south flank of the jet.
9)  Part of the revived SEB became overlaid with orange, methane-bright haze.

______________________________________________________________________

# Introduction

The SEB Revival is the grandest meteorological phenomenon to be seen on Jupiter [ref.1], but only occurs at irregular intervals. Since the modern era of hi-res imaging began, observers and scientists have looked forward to an opportunity to investigate one in detail. The Revival that began in 2010 has been better observed than any before it.

As described in our previous paper [ref.2], sometimes the usual large-scale convective activity in the SEB ceases, the disturbances on the SEBs jet disappear, and the belt begins to brighten. This is a SEB Fade, and is inevitably terminated by a SEB Revival. The Revival always begins at a single longitude, which becomes a persistent source region, and three 'branches' stream out from it in the central, southern, and northern parts of the SEB [refs.1 & 3]. The southern and northern branches follow the rapid jets which bound the SEB: the SEBs retrograde jet (westwards) *[see Footnote 1]*, and the SEBn prograde jet (eastwards). The typical pattern has been described [refs.1 & 3] from the 14 SEB Revivals from 1919 to 1990, and the following introduction is largely quoted from [ref.1, Chapter 10.3].

---

***Footnote 1: The SEBs jet***
 The SEBs jet, with a peak at ~19.5ºS, is the fastest retrograde jet on the planet. The peak speed is up to DL2 ~ +133 deg/month according to spacecraft and ~+120 deg/month for easily visible spots. These dark jetstream spots were initially recorded by visual observers during SEB Revivals, and then during episodes of long-running normal activity. Images from Voyager and Cassini during normal activity showed that they were anticyclonic vortices [ref.1].
 The zonal wind profiles (ZWPs) for the SEBs jet from four spacecraft were compared in [ref.20]. The Cassini ZWP [ref.30] generally appears to be the most reliable; it had a peak speed of DL2 = +128 deg/month (u = -62.0 m/s) at 19.7ºS. ZWPs from Voyager and Hubble gave lower peak values that may not represent the true peak wind speed. However, even the Cassini value may have been influenced by small spots which did not quite attain peak wind speed, and the maximum speed from New Horizons (DL2 = +133.4 deg/month, u = -64.5 m/s, at 19.5ºS: ref.31), despite being the fastest and derived from fewest data points, may in fact be the best estimate of the actual jet peak [ref.6]. (An independent analysis of the same images by Grischa Hahn confirmed a speed of ~+130 deg/mth in some sectors [ref.32].) Our ground-based feature-tracking from 2005 to 2015 supports a peak speed of DL2 = +133 and possibly a broader underlying jet, as shown in [ref.20 Fig.7] and **Fig.10** herein. It is possible that the underlying jet is invariant, but the measured ZWP appears truncated to various degrees, due to crowding by vortices or masking by slower cloud layers.

---

 The Revival always begins when a very dark spot or streak appears across the latitudes of the SEB. Often it is associated with a brilliant white spot on its p. side; sometimes the white spot has been recorded a few days before the streak, in the southern half of the SEB (13-17ºS). It is possible that the white spot always appears first, but visual observers were less sensitive to bright than to dark features. There is usually no visible precursor at the source longitude, but two Revivals appeared adjacent to small dark slow-moving spots on SEB(S).
 From the source point, intense disturbance and/or dark belt material then spreads from the source in three currents or 'branches'.
 *(1) The central branch,* which forms an expanding belt segment whose p. end prograedes along the SEBZ while the f. end remains stationary at the original source longitude. Sometimes this appears to be simply a dark belt segment, often starting off with a bluish-grey colour. But in many Revivals it is very turbulent, with brilliant spots and dark 'columns' continuing to appear for several months at the original source, or at lower longitudes. It is possible that higher resolution would reveal vigorous turbulence in it every time.

The source of an outbreak often remains identifiable for several weeks, and is almost always stationary (mean DL2 = +0.7 deg/month, 1943-1964; but -8 to -20 in 1971 and 1975). The leading edge of the central branch has initial DL2 ranging from -27 to -86 (average, -54); but it may accelerate or fluctuate in speed. The central branch spots (excluding the source) generally move with DL2 between -5 and -70 (average, -33).

*(2) The southern branch*, in the retrograding SEBs jet, usually consists mainly of dark spots, which may be small dark spots like those on other jets, or may be larger. They typically run in chains with a spacing of 12-20°. The average speed has been DL2 = +116 deg/month, which is very close to the overall average for spots tracked in the SEBs jet at all times. The most extreme mean speeds have been DL2 = +89 and +151, for spots from two separate sources in 1943. Often, though, the S branch becomes so turbulent that few spots can be tracked.

*(3) The northern branch* is variable: sometimes it contains the most spectacular spots of all, but sometimes it does not appear at all until 1-2 months into the Revival. Sometimes the spots are visibly deflected northwards and/or accelerated as they pass the Red Spot Hollow, and they may spread across part of the EZ. The northern branch spots may move at any speed between DL2 = -140 and -260 (average DL2 = -197, DL1 = +32). There may be a great range within a single Revival. Visible disturbance tends to persist longer in the N branch than in the other two branches, sometimes being recognisable in the following apparition.

Vigorous new eruptions sometimes occur close to the leading edge of the northern or the central branch. Sometimes a second source appears just like the first; in 1975 there were four such sources. The spots in the Revival can include some of the brightest and some of the darkest that ever occur on the planet. The reviving SEB is often so turbulent that it is difficult or impossible to recognise spots after only 2 or 3 days. There can also be a mixture of strong colours.

After the northern and southern branches have overlapped, the SEB is essentially restored, but disturbances may continue for several months; new spots and rifts continue appearing close to the original source, or at other longitudes.

When the S branch hits the GRS, the GRS almost always fades, beginning within a few days or weeks. Sometimes the dark jetstream spots can be seen running around the curve of the RSH. The breakup of the red material of the GRS is often patchy, and it often recovers temporarily. In any case, within a few months the GRS becomes a light oval with a dark ring around it (11/14 occasions). The dark ring appears to be derived from SEB material.

The final stage of the Revival is an 'orange flush' that spreads across the revived SEB (10/14 occasions). Sometimes there is also darkening or coloration of parts of the STropZ in the aftermath of the Revival, mainly p. the GRS. This may be reddish or yellow coloration, probably extending from the SEB colour; or it may consist of grey shadings or a dark grey S. Tropical Dislocation or Band, possibly derived from grey SEBs jetstream material.

Some Revivals are clearly more vigorous than others, and the relative strengths of the three branches also vary. The central branch can be dominated by one dark belt segment, or by chains of white spots (1949), or by intense turbulence (1928, 1975).

All these observations led to the following tentative model for SEB outbreaks. While the SEB is whitened, an unstable situation builds up below the cloud layer at all longitudes, and can only be released in a classical Revival. Then the eruption begins with a billowing white cloud which, if a Voyager observation of a smaller-scale mid-SEB outbreak is representative, may appear in one of the small cyclonic spots at 16-17°S. The classic dark 'bridge' (streak) across the SEBZ forms at the border of this white cloud. From this time onwards, the outbreak becomes self-sustaining, fixed in the S. Tropical Current, but creating the three spreading branches as it destabilises the neighbouring jetstreams and pours masses of clouds into them.

*More recent Revivals:*
Since the description above, there have been SEB Revivals in 1993 [refs.4 & 5] and 2007 [ref.6]. The Revival in 2007 was well observed and taught us much [ref.6]. Nevertheless, it was weaker than some historical Revivals, occurring with the SEB only partially faded. Also, as the 2007 Revival started close to the GRS, the northern branch did not develop, the central branch did not last long, and the southern branch displayed unusual phenomena because of the presence of S. Tropical Disturbances. Therefore, planetary scientists were still keen to observe another SEB Revival which might reveal the typical phenomena more decisively.

In 2010, the SEB had faded again, much more completely [ref.2]. The Revival began with the planet on the celestial equator so that both southern and northern hemisphere observers were able to image it at high resolution. And the Revival produced brilliant and well-organised spots at the source, and extensive northern and southern branches, so it was typical in every respect. Thus we had the opportunity to study a full-scale example under favourable conditions for the first time. The major phenomena described above have been confirmed by the observations this year, which have also shown that the source and central branch are even more organised than might have been suspected.

The initial outbreak on 2010 Nov.9 was reported in the Journal [ref.7]. We then posted on-line many interim reports on the event [ref.8], including compilations of images of the source region and central branch (on every available rotation up to Jan.9), and of the interaction with the GRS (Jan.11-31). (For a more systematic assemblage of images of the SEB Revival, see the beautiful compilation by Yuichi Iga [ref.9].) Complete reports were then posted with detailed analysis of the whole Revival: reports nos.21, 22, and 24 [ref.8]. This article is largely a distillation of the material in those on-line reports, which may still be consulted for more detailed information ('Supplementary online material'). A short summary has been published [ref.10].

## **Methods of Observation and Analysis**

This report is based on images by numerous observers around the world, as listed on the BAA Jupiter Section web site [ref.8: Report no.25] and on the JUPOS web site [ref.11]. We also used some images posted on the ALPO-Japan web site [ref.12].

Opposition had been on 2010 Sep.21, and although the SEB Revival began 7 weeks later, on Nov.9, it was covered in great detail. In the first month after its appearance, the source region was imaged on 53 out of 73 jovian rotations; up to Dec.25, on 77/111 rotations. Imaging continued with decreasing frequency and resolution thereafter, the last images being obtained in early March, 2011.

Images were almost all taken using webcam selective stacking technology, which has continued to improve in recent years so by 2010, images of unprecedented resolution were being obtained [ref.2]. As usual, most images were produced in visible colour (RGB), some in red or near-infrared (~740-900 nm). Some observers also used a narrow-band filter at 889 nm to produce images in the methane absorption band, in which high-altitude clouds or hazes appear bright.

The analysis is based on work by the JUPOS team [ref.11]: Gianluigi Adamoli, Michel Jacquesson, Hans-Jörg Mettig, and Marco Vedovato. WinJUPOS was created by Grischa Hahn [refs.11 & 13]. The team measured the images using WinJUPOS as usual, and produced charts of longitude vs time for spots in all latitude ranges. For 2010/11, they made 105,118 measurements over the whole planet. However, the raw JUPOS charts gave only a partial view of the SEB(S) activity, as the spots were too complex and numerous to be measured comprehensively. Therefore we also identified major persistent spots visually on image compilations, and combined existing JUPOS measurements of these with manual measurements on additional images, to produce detailed tracks for individual spots.

Throughout this report, drift rates are given in degrees per 30 days in longitude system 2 (DL2). P. = preceding = planetary east (left in images); F. = following = west (right). South is up in all figures.

## Results

**1. The initial outbreak, source, & central branch**

The Revival began with a bright plume erupting inside a cyclonic oval, called 'barge b2', which had been very dark a year earlier, but turned white in summer 2010 [ref.2]. Being cyclonic, the whitened barge was not methane-bright. It was still quiet on Nov.8 (Sadegh Ghomizadeh, in Iran) (**Fig.1**). But on 2010 Nov.9, Christopher Go (in the Philippines) noticed a tiny bright spot within barge b2, at L2 = 288, and immediately alerted the community by email to the possible onset of the Revival. (He had likewise discovered the outbreak of the 2007 SEB Revival, also as a white spot within a pre-existing small barge.)

Ten to twelve hours after its discovery, this white spot (WS1) was already spectacularly brighter (**Fig.1**). In images by Don Parker the new spot was the brightest feature on the planet in all wavebands – near-infrared, ultraviolet, and methane absorption. It appeared to be one of the brightest spots ever recorded in the methane band [ref.7], showing it to be a convective plume of cloud reaching to very high altitude. Indeed it was already discernible in a methane-band image taken in poor seeing by A. Yamazaki (Japan) on Nov.9 (**Fig.1**).

This was the first of 16 such plumes – methane-bright white spots -- that would appear in the expanding disturbance over the next 8 weeks, at a mean frequency of one every 3.7 days. It became evident that these plumes were appearing in just two areas, marking the boundaries of the expanding central branch: its source (f. end), and its leading edge (p. end).

Data on all these white spots are listed in **Table 1**, and figures show the successive plumes (**Figs.2 & 3**) and the larger cells which formed around them (**Fig.4**). The numbering system is the same as in our interim reports, as it is chronological and convenient, even if not entirely consistent.

*Plumes at the source*:
The spots at the source were: WS-1,2,3,7,10,12,13; and we include WS8 and WS11, both ~11° p. the source longitude. As **Fig.5A** shows, the first three spots at the source were still on the extrapolated longitude track of barge b2, and at essentially the same latitude (17.3ºS; barge b2 was at 16.8ºS before the outbreak). As the Revival developed, later spots at the source (WS7, 10, 12) had slightly lower longitude and latitude compared to the track of the barge, but were close to the extrapolated initial track of WS3.

Some plumes at the source grew or brightened very rapidly even within 10-20 hours, especially WS1 and WS3. However, plumes at the source did not appear to expand much thereafter. (Plumes only expanded greatly when they drifted north of ~13.5ºS, notably WS-N and WS9a at the front; see below.) But if the plume itself remained small, a large expanding 'cell' always formed around it, bounded by dark streaks.

Each new plume at the source, by the second or third day (**Fig.2**), had acquired a very dark spot or streak bordering its f. side. Over subsequent days, this extended north to form a very dark narrow streak – the classical 'dark column' that has often been the most conspicuous feature of the source in historical SEB Revivals. This streak defined the f. boundary of the cell surrounding the plume, and persisted even if the plume, being smaller than the cell, was not in contact with the streak, or if the plume faded away altogether (**Fig.4**).

These cells were long-lived and comprised the whole of the central branch (**Fig.4**). WS3 prograded and merged with or replaced WS2, but maintained its cell around it. The pattern was also maintained by the only two bright spots that appeared within the central branch rather than at its p. or f. edges: WS8 appeared within a space vacated by the rapid motion of WS3, temporarily forming a new cell f. it; and WS11 formed within the cell of WS7, after WS7 itself had disappeared. The evidence suggests that WS8 and WS11 were weaker than other plumes: WS8 only lasted for 3 days (during which there were no methane images), and WS11 was initially not methane-bright, although it became so later.

WS3, and subsequent plumes at the source, after being near-stationary for a few days, suddenly accelerated p. (eastwards) (**Fig.5**). Remarkably, this happened without any sudden change in latitude, although each of them was gradually drifting north, so the mean latitude during their rapidly prograding phase was 14.2 (±0.1) ºS. Nor did the accelerations coincide with any obvious change in appearance: each plume was still a compact, very bright spot, although some faded from high to moderate methane-brightness around this time. The prograding speeds ranged from $DL2 = -25$ to $-53$ deg/month, with a mean of $-40$ deg/month (for 5 plumes). These were the speeds of the mature cells constituting the central branch.

Of the later plumes, WS10 was outstandingly methane-bright like the earlier ones, but WS7, WS8 and WS11 were either not methane-bright or only briefly so, and WS12 was only weakly methane-bright. Thus the frequency and scale of the plumes at the source seems to have decreased after mid-Dec. And after Jan.5, the pattern of cells was no longer evident, as there was greater small-scale complexity throughout the central branch without very conspicuous spots. A final small bright spot appeared at the source on Jan.14, but failed to grow, and was not visible after Jan.15.

It may be significant that a bright plume appeared somewhere in the disturbance every 2-6 days up to Dec.13 (excluding WS9; see below). This overall rate is more regular than the rates for either the source region or the front taken separately (**Table 1**). Moreover, sometimes one bright plume appeared around the time that a nearby one faded or disappeared (WS1, WS-N; WS7, WS8, WS10, WS11). This suggests that the disturbance tended to focus its convective activity into just one new plume at a time. WS9 may be 'the exception that proves the rule' as by the time it appeared, the front was sufficiently far from the source region that correlation was no longer maintained. However, methane images did sometimes show several plumes active simultaneously.

*Plumes at the leading edge*:
The first of these appeared just 3 days into the outbreak: WS-N, on the Np. edge of WS1. As subsequent weeks passed, it became evident that such plumes were appearing along a broad front comprising the oblique leading edge of the advancing disturbance: 'the mother of all weather fronts', with vigorous bright storms erupting at any point along it from 14ºS to 19ºS. By late Nov., this leading front was well defined (**Fig.4B**), consisting of the extended bright area derived from WS1, in which the new bright plumes arose, and an oblique dark streak on its f. side derived from the original column f. WS1. After WS-N the plumes were WS4, WS5, and WS6, at a range of latitudes from 16ºS to 19ºS, but all with $DL2 \sim -3$ to $-6$. WS6 moved south while WS5 moved north in a cyclonic swirl (Dec.1-3). There were other, smaller and shorter-lived bright spots in the same latitude range from Nov.27 onwards, including a chain of five on Dec.15, one of which became very bright, along the S side of the expanding central branch. Later, there appeared the very conspicuous WS9 and WS9a; and several unnumbered spots at the start of January, again spanning a great range of latitudes (**Figs.2 & 5; Table 1**). All these spots were methane-bright, most of them being as brilliant as the plumes at the source.

Both WS-N and WS9 in turn extended northwards and then eastwards, still methane-bright, each becoming a long oval (surrounded by a large dark loop), which accelerated into the northern branch of the Revival (see below).

*Latitudes and speeds:*
The chart (**Fig.6**) shows the range of speeds and latitudes of the plumes, and compares them with the zonal wind profiles (ZWPs) deduced from spacecraft images. The first three plumes at the source all appeared on the same track as former barge b2, and in the same latitude (17.3 ±0.2 ºS), which lies on the spacecraft ZWP. But with subsequent northward shifts of these spots, and others appearing in other latitudes, a very wide scatter of latitudes and speeds developed. For plumes with slow speeds, their broad range of latitudes confirms that the cells were coherent structures across the width of the SEB, from 19ºS to 12ºS. White spots at the p. or f. edges of the central branch, with DL2 between +4 and -12 deg/mth, were found all across this latitude range, including WS2 and WS3 which descended by several degrees in latitude without changing their speed. On reaching 14-15ºS, most spots did accelerate, but to a very variable extent, with speeds ranging between -6 and -53 deg/mth.

Mean speeds for white spots, from **Table 1**, with standard deviations, were:
--for latitudes 18.0 to 15.5ºS:  mean DL2 = -4.3 (±5.6) deg/month.
--for latitudes 15.5 to 12.5ºS:  mean DL2 = -33.3 (±15.9) deg/month.

*The source and central branch, mid-Jan. to mid-Feb.:*
In 2011 Jan. and early Feb., the overall structure of the central branch continued as before (**Figs.4 & 7).** Thus there were new bright spots still appearing in the source region near L2 ~300; and very complex disturbance p. this; and an oblique leading edge of the central branch, prograding towards the GRS. At the source, further small brilliant white spots appeared on:
Jan.3/4 (L2 = 293; WS13; methane-bright on Jan.9);
Jan.14 (L2 = 307; but failed to grow, and was not visible after Jan.15); and
Feb.5 (L2 = 305; also very methane-bright; disappeared by Feb.8).
This list may be incomplete because observations were sparser at this late stage of the apparition; however, the plumes were certainly less frequent and less persistent than before.

     Starting in early Dec., each cell developed reddish-brown diffuse shading in the southern part. In Jan-Feb., the central branch was too complex for tracking individual spots, and showed extensive reddish-brown colour.

     The leading edge prograded with mean DL2 ~ -32 deg/month (Nov.21-Feb.18), but it was irregular in shape (often being disrupted by the bright spots) and in motion. It was roughly stationary in L2 in early Dec. and mid/late Jan., but had DL2 ~ -42 deg/month at other times.

     The leading edge reached the Red Spot Hollow (RSH) about Feb.20 (**Fig.13**). It was not obvious whether this caused any change, as the RSH was already much darkened by material from the northern branch (see below).

*Source and central branch:  Physical implications*
The remarkable brightness of the initial spot in the methane band implied that it was a cloud plume projecting above the normal cloud-tops, exceptionally high and/or dense – presumably a convective plume which had burst up through the clouds from a source below. This also applies to the subsequent bright plumes, although the later ones were not as bright as the first ones. The plume is thought to represent moist convection from the water cloud layer [refs.3 & 14], i.e., it is likely to be an exceptionally large version of the bright spots that appear in the SEBZ in normal times, which are giant thunderstorms [ref.15].

     The appearance within a barge was significant, as the 2007 Revival also began in a barge [ref.6, especially report no.6 therein], and Voyager images in 1979 had shown a mid-SEB

outbreak beginning precisely in the centre of a miniature barge at 16.3ºS – a presumed cyclonic circulation [refs.1 & 16]. Why would this be? Although the barges had been filled in with white cloud in 2010, their visible outlines implied that they retained cyclonic vorticity, and the fact that one still appeared 'warm' at 10.8 um wavelength [ref.17] suggested that they might still be warmer than their surroundings deeper down. It remains to be determined whether either of these properties could make the barge a favoured site for the onset or growth of massive moist convection.

The barge itself could not be observed once the outbreak had started, so we cannot tell whether its cyclonic circulation was maintained, or whether it was totally overwhelmed by the powerful convective storm that then established itself.

In addition, white, methane-bright plumes were observed to erupt along the leading edge of the central branch. In past revivals, the leading edge of the central branch has often been identified as a coherent structure, often with a bright white spot there [ref.1]. The hi-res observations in 2010-11 strikingly confirm that this was a distinct structure and a locus of extremely vigorous storms, comparable to those at the source; and that there was little new activity in between these p. and f. ends of the central branch.

This year's observations also revealed an unexpected cellular structure in the expanding central branch. Lower-resolution observations from previous SEB Revivals suggested that new bright spots would typically expand to span the belt (as they do during normal mid-SEB activity); but in 2010, most plumes at the source did not appear to expand much, whereas a large expanding 'cell' always formed around the plume, drifting much more slowly than the ZWP would predict; these would have been regarded as white spots in lower-resolution observations. Thus the developing Revival behaved as a coherent whole. Bright spots tended to remain in their cells, with little sign of the normal gradient of speed across the belt, except for those spots and streaks which broke free into the northern or southern branches.

Questions remain as to how abnormal this behaviour is. In normal times, there are frequent bright spots arising and circulating within 'rifted' sectors of the SEB. It is not yet clear whether these are as coherent as the cells identified in the SEB Revival, and they are evidently sheared by a wind gradient across the belt as shown in ZWPs. However, ZWPs of the SEB from spacecraft show broad ranges, sometimes multiple gradients (**Fig.6**), whose origins have not been reported. Unpublished assessment of ZWPs for the SEB in separate longitude sectors, in the 1990s from HST [ref.18], and in 2012 both from HST and from amateur images [ref.19], shows that the slower ZWPs were from sectors that included the GRS and rifted region f. it, while the faster ZWPs (~25 m/s faster in the prograding direction from ~12-14ºS) were from undisturbed sectors. This suggests that expanding plumes or cells in the rifted sector in normal times may retain slower drifts influenced by their latitude of origin. This behaviour resembles, partially, the behaviour of the plumes and cells in the 2010 SEB Revival. But a detailed study of this topic remains to be done.

Each cell in the SEB Revival eventually developed reddish-brown diffuse shading in its southern part. These amorphous coloured regions were light in methane images, indicating that the brown material was a high-altitude haze, not a thinning of the cloud. It may have been emitted around the plumes, as reddish haze tends to appear at locations of vigorous activity on Jupiter. It presumably corresponds to the typical 'orange flush' that often (though not always) develops over the SEB after a Revival.

**2. The southern branch**

The southern branch comprises the dark spots which move rapidly westwards with the strong SEBs jet. (For this section only, 'faster' and 'slower' refer to speeds in the retrograde

(westward) direction, i.e towards increasing longitude.) According to the New Horizons profile, the peak speed is DL2 ~ +133 deg/mth at 19.5°S (see Introduction & Footnote 1).

*The pre-existing chain:*
Before the Revival, and in sectors not yet affected by it, there was a regular chain of small bright spots 5-7° apart, separated by faint grey patches (also called 'projections' from the residual faint SEBs), all on the south flank of the SEBs jet. They were moving much more slowly than the usual jet speed. The mean speed was DL2 = +69 deg/mth (ranging from +80 to +51 deg/mth in different sectors), and the mean latitude for the spots was 20.7 (±0.25) °S. This chain was fully described in Paper I [ref.2], and we have shown [ref.20] that it was a manifestation of a wave-train along the peak of the SEBs jet at 19.5--20°S, with a phase speed much less than the wind speed of the jet.

In addition, there were several short gaps in the chain which moved with DL2 = +134 deg/mth, i.e. the peak speed of the jet, indicating that the jet wind speed still existed at some level, even though no distinct features could be seen on it.

*Southern branch of the Revival:*
The longitude-vs-time chart for the dark spots in the southern branch is shown in **Figs.8 & 9**, and some images are shown in **Figs.7 & 9 & 11**. The early spots are still labelled by their original designations: P1 to P3 were the first, rather small and short-lived spots, then DS1 to DS6 were larger dark spots. All the spots present after mid-Dec. have been re-labelled alphabetically from A to T.

*[Small type for the following details:]*

The southern branch began on Nov.14, when a faint projection in the pre-existing chain (P1) darkened as it passed the source. It suddenly accelerated to full jet speed (DL2 = +132, from Nov.16-25, then slackening to +110; **Fig.9**), without decrease in latitude; and split into two (P1, P2).

More substantial dark spots began with DS1, forming on Nov.20-21 when a new, very dark spot at the source extended Sf. to darken and accelerate the next projection in the chain. Although the pre-existing projections could not be tracked thereafter as they passed the disturbed source region, the morphology and the JUPOS chart (**Fig.9**) suggest that subsequent dark spots in the southern branch developed in the same way. Thus DS2 to DS5 formed from Nov.24-27, all with DL2 ~ +108 to +120. They never adopted the oval form of vortices. Their shapes and arrangements were changing very rapidly; however some of the best images showed they were separated by and curled around small white ovals at ~20.1°S, separated by 7-9° – possibly the same ovals that formed the pre-outbreak chain. More very dark spots continued to emerge from the chaotic region Sf. the source, of which DS6 was the most substantial.

However, by mid-Dec., most of the dark spots had faded, leaving only three with full jet speed: P1 (rapidly fading), DS3 (which had become very large and dark, spanning the whole SEB(S)), and DS6 (still dark). DS1 and DS5 also survived but had become concentrated on the SEB(S) south edge and accordingly decelerated, sliding past DS3 and DS6 respectively.

Following (west of) DS1 and DS3, as most of the leading dark spots disappeared or merged, the previous faint chain pattern re-established itself, with DL2 ~ +70 as before. But spot P1 still persisted with its faster speed, DL2 = +110, although it looked much like the other projections except that it was darker. Detailed analysis of images on Dec.11-14 (**Fig.9B**) shows that P1 was a southerly patch that actually shifted from one slower-moving projection to the next. It disappeared around Dec.24, leaving only the pre-existing chain.

Other sectors within the reviving SEB(S) again developed chains of small bright ovals (at 20.1°S) which resembled the pre-outbreak chain and had similar slow speeds (see below). The edges of the revived SEB(S) were at 19.4 and 21.6 (±0.3)°S. Thus, its north edge (*sic*) was close to the canonical latitude of the SEBs jet peak, and the new belt encompassed the latitude of the pre-outbreak chain (**Figs.7&11**).

In late Dec., the sector from the source to DS5, perforated by small white ovals, generated another series of very dark spots, including small dark spots in mid-SEB(S), and large dark streaks on the south edge. Although they were conspicuous, they did not show long steady tracks, possibly because the large spots were actually transient assemblies of central and southerly spots with different speeds. In January, DS5 (now labelled F) merged with spots p. and f. it (E,G,H), forming one very dark spot on the S edge ('HE'), with the same drift as the adjacent white ovals in mid-SEB(S) (**Fig.8**). Continuing to emerge closer to the source were more small dark spots, some of which became southerly and thus had slower (but variable) retrograding drifts. After mid-Jan., no faster-retrograding spots remained, while slower-retrograding speeds developed, so drifts varied from +70 deg/mth to ~0, in a very complex pattern (**Figs.8 & 11**). Among the last dark spots produced in the southern branch were two very conspicuous and southerly ones, P and T, which soon decelerated and remained quite close to the source. Spot P was initially a very dark oval, later a ring with light (reddish?) interior, and was the only dark spot which looked like a vortex.

The drift rates and latitudes for the southern-branch spots, over the whole Revival, are summarised in **Table 2 and Fig.10d**. Most of the major dark spots at 20.0 to 21.3ºS showed speeds in the jet peak range (+114 to +134 deg/mth). All the more southerly dark spots (21.4 to 21.9ºS) moved more slowly, most of them with $DL2 \sim +44$ to +60, which is consistent with the spacecraft ZWP. Meanwhile, small northerly spots (20.0ºS) also had a slow speed, $DL2 \sim +51$, along with the small white ovals that were forming a renewed slow-moving chain within the SEB(S).

Although the SEB(S) showed a degree of darkening up to the Red Spot Hollow (RSH) in early Jan., only one substantial dark spot ever reached the RSH; this was DS3(=A), which arrived at the RSH on Jan.17 (see below). Other dark spots in the SEB(S) tended to fade away as they proceeded to high longitudes. Nevertheless, from mid-Jan., the SEB(S) was fully revived from the source to the RSH (**Fig.7**). The SEBZ alongside it was still bright white, unaffected by the Revival.

*The slow-moving chain of white ovals in the revived SEB(S):*
We earlier reported speeds for the pre-existing chain in Nov. at longitudes not yet affected by the Revival: mean $DL2 = +71$ (Nov.1-25) [ref.2]. This speed was maintained in Dec. for the same sector ('Sector 0' in **Figs.7 & 8 & Table 2**), which remained ahead of the Revival.

Chains of white ovals also re-established themselves in sectors of the revived SEB(S) (marked by red circles in **Fig.11**). Sector 1 extended from DS3(A) back to the main dark complex (HE), in Jan., where other substantial dark spots faded away and the chain of white ovals reappeared with $DL2 = +43$ to +57 (mean +51; lat.20.5ºS). Sector 2 was alongside or within the major dark features (M, HE, etc.). Here too, white ovals were reappearing, with essentially the same speed and latitude as in Sector 1 ($DL2 = +44$ to +55; lat.20.3ºS). Thus, the revived chain of ovals had the original slowly-retrograding speed – half the speed of the normal jetstream and of the early dark spots of the Revival, even though they were in the same latitude!

### *Southern branch: Physical implications*

*What were the chains of ovals?*
We have shown evidence that these chains, both before and after the outbreak, were wave-trains, with a phase speed about half the wind speed for that latitude [ref.20]. Although the white ovals may have been at a higher altitude than the usual cloud-tops to which the ZWP refers, the evidence in this and subsequent years is that the wave-train itself is embedded in the observed peak of the SEBs jet at cloud-top level [ref.20].

*What were the dark spots?*
In this Revival, as in most previous ones, the reviving SEB(S) appeared highly disturbed, and we expected that the southern branch would consist of vortices and turbulence as intense as in the Voyager or Cassini images [ref.1]. However, the dark spots formed in the outbreak were apparently not vortices, but irregular dark patches, streaming into the SEB(S) from the source. And they did not cause lasting disruption to the pre-existing wave-train of white ovals. The first dark spots appeared to be 'projections' between the white ovals that filled up with dark material flowing from very dark spot(s) at the source, and most of them soon faded again. Later SEB(S) spots, being larger and darker, appeared to be amorphous dark streaks and patches, but often appeared to be moulded around the white ovals, and ended up at a southerly latitude with slow drifts, in accordance with the ZWP (**Fig.10d**). These slow drifts exactly matched those of the wave-generated white ovals alongside.  Therefore, the dark spots were  much less disruptive than expected.  We can suggest two possible explanations, which are not mutually exclusive.

    (i) They may be dark aerosols, as has recently been proposed for some spots on Saturn [ref.22].  Cassini images revealed dark grey spots forming in the wake of Saturn's persistent thunderstorms, and the dark material could be carbon (soot) generated by the lightning. If this process also occurs on Jupiter, given that the bright SEB plumes are probably intense thunderstorms, the retrograding dark spots may contain soot generated therein, distributed by the winds and waves of the SEBs jet.

    (ii) The dark spots may be clearings in the mid-level clouds, produced by pulses of warm air emitted from the source, but without turbulence so they did not disrupt the wave-generated ovals of white cloud at higher level.  Thus, they only appeared dark while passing between the white ovals.

    Whether either or both of these hypotheses is true might be determined from professional infrared observations.

*Were there any changes to the zonal wind profile during the Fade and Revival of the SEB?*
As noted above, the slow speed of the chain of ovals was a wave phase speed, not a change in the ZWP [ref.20].  The mean speed of the more rapidly retrograding spots in this Revival (DL2 = +117 deg/mth: **Tables 2 & 3**) is typical of SEB Revivals [ref.1; see Introduction], and some of these spots were at ~20ºS, also consistent with the usual jet peak.  However, the dark spots with peak jet speed were spread over a much broader latitude range than would be expected from the spacecraft ZWPs.

    This anomaly is  not limited to the SEB Revival.  We have found that in normal times, and in the early stages of the 2007 SEB Revival, vortices move with almost peak jet speed up to ~1.5º south of the jet peak in spacecraft ZWPs **[Fig.10a-c]**.  This anomaly was even observed in the Cassini data, so it must coexist with the Cassini ZWP.  It could perhaps be an aspect of the vortex dynamics, as vortices do not necessarily respect the ZWP [ref.21].  On the SEBs they move at a speed more appropriate for their north edge (on the jet) rather than for their centre. The vortices are clearly at cloud-top level, and appear to arise by interactions of cloud-top currents according to spacecraft images [refs.1 & 15].

    However, the new, rapidly retrograding spots that appeared in the Revival likewise lay further south, forming a broader zonal drift profile that coincided with that of the vortices in other years, although the 2010 spots did not appear to be vortices.  Thus, they were inconsistent with the known ZWP.  Three alternative hypothetical explanations can be suggested:

    (i) The ZWP changed immediately when the SEB Revival started, broadening the jet peak to the south.  This would be consistent with the behaviour in the 2007 Revival when only the earliest southern-branch dark spots were anomalously far south. It is conceivable that the SEBs jet broadened at depth, either as part of the build-up to the outbreak, or as an immediate reaction to it. Possibly the outbreak source was such a large, deep and powerful storm that it caused a sudden massive perturbation of the jet. This would be the first time that a definite increase in zonal winds has been demonstrated by modern observations [*Footnote 2*].

*Footnote 2*: The SEBs has retrograded even faster in the past, which I have attributed to S.Tropical Disturbances [ref. 1]. In other cases where a jet has apparently increased in speed – the NEBs and NTBs – this is now thought to be due to a permanent deep super-fast jet emerging at the surface.  That is not the case here: the peak jet speed has not increased, but has broadened to the south.

(ii) The dark spots were at a deeper level, where the ZWP is different in that the jet is always broader than at the cloud-tops. (This broader profile, as sketched by a grey line in **Fig.10**, would be more compatible with the barotropic stability criterion than the sharp peak of the ZWP [refs.23 & 24].)  As suggested above, the dark spots were clearings in the clouds, but without turbulence so they did not disrupt the overlying white ovals of the wave-train.

(iii) As (ii), but the dark spots were deep-seated vortices, which did not penetrate fully to cloud-top level, so their vortical shapes were never observed as they passed below the white ovals.  In this case, there was no difference in the ZWP, and the speed discrepancy was due to the vortex dynamics as proposed above.

## 3.  The northern branch

Initially the outbreak had little effect on the SEB(N).  However, a blue-grey streak or wedge repeatedly developed p. the leading edge of the central branch (**Fig.4**, & Refs.8&9).  It first developed from Nov.14 onwards, from a small blue-grey spot p. WS-N which elongated rapidly eastwards.  By Nov.22-24 it was faint visually, but in subsequent weeks a distinct blue-grey streak reappeared several times and repeated the process. Although the visible shading was variable, it became a persistent methane-dark feature.

Otherwise, the northern branch up to mid-Dec. comprised just a narrow dark segment of SEB(N), which reached the GRS on Dec.11, and continued past it.  This part p. the GRS was dark brown, but further f. it still had the mysterious slightly greenish tint of the faded SEB(N), giving way to dark bluish or grey in the oblique streaks at the p. end of the central branch.

On Dec.11, WS9 was eddying clockwise, scooping the SEB(N) into a very dark grey (and methane-dark) streak on its p. and S sides (Dec.13-23) (**Figs.4&7**). As WS9 elongated rapidly eastwards, the dark streak elongated even faster to become the first substantial feature in the northern branch.  By Jan.1, it had become a long, indistinct, dark (and methane-dark) sector of revived SEB(N), with some small dark spots prograding along it.  Also, a fainter blue-grey wedge  reappeared on its S edge, p. the still-bright leading edge of the central branch.

On Jan.5, there was a dramatic change: this large, pale blue-grey wedge suddenly broke up into an impressive series of  'waves' **(Fig.12A)**.  This occurred within only 20 hours!  Over the next few days, the waves became much darker and grey. They showed a range of speeds from DL2 ~ -150 to DL2 ~ -90 **(Fig.12B)**, so that their separation increased from 6º (Jan.5) to 9º (Jan.12), as they approached the GRS.

As they prograded past the GRS, successive spots formed a large, dark,  almost black spot on its N edge. (The same appearance featured in the 1990 SEB Revival.)  This spot appeared to be roughly fixed there, but this was an illusion due to successive spots becoming very dark as they passed through this position **(Fig.13)**. The tracks of these spots were often disturbed as they passed the GRS, but did not show systematic change of speed **(Fig.12B)**.

After passing the GRS, in Jan.-Feb., the spots had DL2 ~-125, spacing 9º. They were tracked until approximately Feb.10 when the leading edge of the series was near L2 ~ 40.

## 4.  Interaction with the GRS

Both southern and northern branches began to affect the Red Spot Hollow (RSH) tenuously around 2011 Jan.5, darkening its rim and also the SEB(S) f. it **(Fig.13)**.  At this time the incoming southern branch just consisted of slightly darkened projections in the pre-existing chain, ahead of the main dark spots, while the incoming northern branch was the series of prograding dark spots that suddenly formed on Jan.5. Each SEB(N) dark spot in turn became very dark when due N of the GRS, and dark material streaming Sf. from these dark spots was probably the main cause of the darkening of the RSH. But the SEB(N) dark spots did continue prograding past the RSH, with bright bays developing between them **(Fig.12B)**. The previous bright plume N of the GRS had already been displaced by these spots, and lost its identity in mid-Jan. **(Fig.13).**

Meanwhile, two dark spots approached on the STBn jetstream in mid-Jan. **(Fig.13)**: one [a] was deflected N alongside the f. edge of the GRS, but the other [b] continued prograding past the S edge of the GRS then dissipated as a faint brown streak in STB(N). (Before the Revival, many spots had behaved like [b], and one or two like [a]: see our reports nos.8 & 12 in ref.8.) The remnant of [a] contributed to the dark collar developing around the GRS.

In late Jan., this collar was further reinforced by a very dark grey streak emitted from the dark spots N of the GRS, which extended anticlockwise around the GRS. Its leading part emerged at the p. end of the GRS as a little dark streak (Jan.29), which then expanded into a large, light brown loop - a bizarre sight in Don Parker's multispectral images on Jan.31 **(Fig.13)**.

The first large dark spot on SEB(S) (DS3 = A) had arrived at the GRS on Jan.17 **(Fig.13)**, and its dark brown material flowed around the RSH over the next few days, creating a mess at the northernmost point where it met the very dark spot from the SEB(N). There was no indication that DS3 persisted any further; however, material from it may have contributed to the complex streaks and disturbance around the GRS in late Jan. The SEB(S) trailing behind it was generally dark with a cellular structure, but no further distinct spots were tracked from it into the RSH or GRS.

In Feb., similar activity continued with greater intensity. More prograding dark spots on SEB(N) (e.g. one cluster marked by a green arrow in **Fig.13**) were piling into the very dark spot N of the GRS; from there, dark material was streaming Sf. around the rim to another very dark patch at the f. end of the GRS; and from there, streaks and spots were running p. around the S. edge. One very dark spot followed this course, anticlockwise around the GRS to its p. side, from Feb.9 onwards (red arrow in **Fig.13**). It was probably this which emerged on the p. side as a large brown streak in the first week of March (red arrowhead) – although the arrival of the central branch at the RSH around Feb.22 may also have contributed more dark material. This brown streak was only seen in the last, lo-res images of the apparition in March **(Fig.13)**, but it was probably the beginning of a massive S. Tropical Band seen in the next apparition.

## 5. Aftermath in 2011  [ref.25]

*South Tropical Band*
After solar conjunction, images by T. Akutsu in 2011 April-May revealed a long, dark grey S. Tropical Band which was a prominent feature in the following months [ref.25]. By July, this massive dark band almost surrounded the planet. It was similar to one seen in 1991 after the 1990 Revival, and is an example of those diverse dark formations which sometimes develop in the STropZ, prograding, from the p.(E) end of the GRS, usually at the end of an SEB Revival or (in more normal times) after a series of SEBs jetstream spots has disappeared into the GRS rim [ref.1, pp.203-214].

The GRS was a uniformly pale orange oval, completely enclosed by a substantial dark grey rim.

*'Orange flush' and quiescence of the SEB*
The SEB was fully revived from the start of the apparition in 2011 May, although white spots (convective storms) were still arising close to the position of the source of the Revival and prograding in the northern half of the belt. Meanwhile, in June-July, the southern half, for ~60° f. the GRS, was unusually featureless and reddish, a pale orange-brown colour. This was probably the end result of the reddish colour seen in a narrow southerly strip during the Revival, and it may well have been a restricted version of the 'orange flush' that was often recorded visually after previous SEB Revivals [ref.1]. Notably, this sector was very light in methane images [ref.26], consistent with the 'flush' being a high-altitude orange haze overlying the belt.

In mid-Sep., although the orange, methane-bright haze had largely disappeared, the complete quiescence of the SEB (and GRS) suggested that the belt might be about to start fading again. However, on 2011 Sep.21, the first convective white spot since the Revival appeared just f. the GRS, initiating the normal convective activity which has continued since then.

## Discussion: The organisation of the SEB Revival

The parameters of the 2010 SEB Revival are summarised in **Table 3**. The drift rates were all within, or close to, the historical ranges summarised in the Introduction (except for the slow-moving spot-chain or wave-train on SEBs).

The observations analysed here are consistent with previous accounts of SEB Revivals [refs.1 & 3]. But the unprecedented temporal and spatial resolution of the 2010 observations has allowed us to define the features much more precisely, and to infer dynamical processes that were previously obscure. They largely validate the historical descriptions of these events: the major features recognised in them, albeit at lower resolution, are indeed the large structural features described here. The implications for the physical nature of the disturbances have been discussed in previous sections.

Here we list the major conclusions about the 2010 SEB Revival, and ask: Did they also apply to the 2007 Revival [ref.6], and to other Revivals in the historical records for which sufficiently detailed information is available [ref.1]?

1) The Revival started with a bright white spot (plume); the typical very dark 'column' developed soon after.
2007: Yes.
Previous Revivals: Yes, at least sometimes. These observations support our previous speculation [ref.1] that the white spot always appears first, but is less likely to be noticed against the bright zone. In six outbreaks in four Revivals, a bright white spot at 13-17°S was indeed observed 2-5 days before the canonical dark 'column' (1943B, 1949, 1971A,B, 1975A,B). Here, 'A' denotes the primary outbreak, 'B' a secondary outbreak, and the observations in 1971A and 1975A were by professional photography [refs.1,27,28]. Thus, visual observers detected the white spot first in only one primary outbreak, but hi-res imaging has detected it first in most of the recent outbreaks. The more frequent detection preceding secondary outbreaks was probably because visual observers were paying closer attention to the SEB once a primary outbreak had started.

2) The Revival started with a bright white plume erupting in a pre-existing barge. Subsequent white plumes continued to appear on the track of this barge, which was the location of the sub-surface source of the whole Revival.
2007: Yes.
Previous Revivals: Sometimes. Revivals broke out adjacent to small dark slow-moving spots on SEB(S) in 1943, 1949 [ref.1], and 1993 [refs.4&5]. But other Revivals might have started in barges that were too small or faint to be seen, just as the barges in 2007 were very small, and the barges in 2010 had whitened to become invisible long before the Revival started.

Once the outbreak has begun, the source often remains identifiable for several weeks, and is usually near-stationary in L2 (mean DL2 = +0.7, 1943-1964) [ref.1]. This description is consistent with the 2010 source, and with the usual slow drift of mini-barges in the SEB.

3) These plumes were extremely methane-bright (thrusting up to very high altitudes), especially when new.

2007: Yes in HST images of a newly erupting plume on June 5 [ref.29]. But in ground-based methane images (which were adequate though not plentiful in 2007), the white spots were only modestly methane-bright, and one new one observed (May 27) was not methane-bright at all. All this suggests that the plumes were less vigorous in 2007.

Previous Revivals: Yes, on the two occasions when methane images were obtained. Ten days into the 1971 Revival, the white spot at the source was reported as very methane-bright [ref.28]. So was the white spot in 1993, on the second day [ref.3].

4) Brilliant white spots (methane-bright plumes) also appeared along the leading edge of the central branch.

2007: No. We reported [ref.6]: "*Central branch* ...early bright spots were rapidly distorted or disappeared, and later ones were very small and transient, with much small-scale turbulence; and even the leading edge of the ensemble was very oblique." As we thought at the time, the 2007 Revival was less energetic than some historical examples.

Previous Revivals: Yes, sometimes. "Vigorous new eruptions sometimes occur close to the leading edge of the northern or the central branch." [ref.1]

5) The central region of the outbreak was composed of large convective cells. Each cell was initiated by a bright plume, although the plume only occupied a part of the cell, and a very dark streak ('column') persisted to define its f. edge.

2007: Probably, but the cells were perhaps smaller in 2007, because of crowding near the source, and because SEB(N) was already broad and dark and they did not encroach on it.

Previous Revivals: Probably, sometimes. The central branch may appear as simply a dark belt segment; but in many Revivals it is very turbulent, with brilliant spots and dark 'columns' continuing to appear for several months at the original source, or at lower longitudes. The appearance may be consistent with a series of cells, but cannot be shown to be so systematically, as any cellular structure may be either below the limit of visual resolution, or masked by apparently chaotic variability. The one Revival in which cellular structure was distinct was in 1949, when the central banch consisted mainly of a series of bright spots, separated by dark columns or patches, prograding from the source; they were produced at a rate of one every 6 days. [ref.1]

6) The southern branch began with darkening of pre-existing mini-projections.

2007: No, because: (i) as the SEB fading had not proceeded for long, no pattern of mini-projections was present; (ii) as the SEB outbreak was adjacent to a S. Tropical Disturbance, any such coherent features on SEBs would have been diverted away.

Previous Revivals: Unknown, due to insufficient resolution.

7) Dark spots in the southern branch were not coherent vortices.

2007: No: many or all of the SEBs dark spots were probably anticyclonic vortices.

Previous Revivals: Unknown, due to insufficient resolution.

8) Dark spots in the southern branch had typical SEBs jetstream speed but were unusually far south.
2007:  Yes, but only for the early dark spots; later ones were on the usual ZWP.
Previous Revivals:  Unknown, due to insufficient latitude information.

9) Part of the revived SEB became overlaid with orange, methane-bright haze, so in the following year it was ambiguous whether the SEB would resume fading or resume normal activity.
2007:  No; rifting continued and proliferated.
Previous Revivals: Yes; the 'orange flush' was noted after most Revivals [ref.1], although it was restricted to a short sector f. the GRS in 2011.  Colour and methane-band images from HST showed the same aspect f. the GRS in 1991 and 1994, after the Revivals of 1990 and 1993 [ref.26].  The subsequent quiescence of this sector seems to be an under-appreciated aspect of this phase of the SEB cycle.  After each of the Revivals of 1971, 1975, 1990, and 1993, the SEB became quiet in the following year and looked as though it would fade again; on two occasions a Fade indeed proceeded, but on two occasions normal rifting resumed.

These observations have greatly clarified our knowledge of a typical SEB Revival, but have left some important questions for future research:
--Why does the outbreak appear within a cyclonic oval?
--Why is this deep source so localised and powerful?
--What are the dark spots in the southern branch?
--How can they co-exist with the wave pattern on SEBs without destroying it?
--How does the zonal drift profile of the visible disturbances relate to the usual ZWP?

-----------------------------------------------------------------------------------------------------

# Tables

*[The original Tables are given in a separate Excel file. My Excel uses the continental style of commas for decimal points.]*

### Table 1.  Life histories of the bright plumes and cells.

| Table 1 | Life histories of the bright plumes and cells | | | | | | | | |
|---|---|---|---|---|---|---|---|---|---|
| *No.* | *Location* | *First observations:* | | | *Last (w.s.):* | | *Track (cell):* | | |
| | | *Date* | *L2* | *Lat.* | *Date* | | *Dates* | *DL2* | *Lat.* | *SD* |
| WS1 | Source | Nov.9 | 288 | -17,1 | Nov.14 | WS1 | Nov.9-17 | -12 | -17,1 | 0,52 |
| WS-N | Front | Nov.12 | 286 | -14,0 | (-->WS5 ) | WS-N | Nov.12-27 | -17 | -13,6 | 0,69 |
| WS2 | Source | Nov.14 | 291 | -17,5 | Nov.20 | WS2 | Nov.12-21 | -6 | [-17,5 -> -14] | |
| WS3 | Source | Nov.17 | 294,5 | -17,3 | Dec.13 | WS3 | Nov.17-25 | 4 | [-17,2 -> -15,4] | |
| | | | | | | WS3 | Nov.26-31 | -53 | -14,3 | 0,46 |
| | | | | | | WS3 | Dec.2-14 | -11 | -12,1 | 0,53 |
| WS4 | Front | Nov.19 | 290 | -19,4 | Nov.24 | WS4 | Nov.20-29 | -3 | -18,9 | 0,55 |
| WS5 | Front | Nov.23 | 284 | -16,6 | Dec.12 | WS5 | Nov.24-Dec.12 | -6 | -17,4 | 0,62 |
| WS6 | Front | Nov.28 | 279 | -16,1 | Dec.8 | WS6 | Nov.28-Dec.8 | -6 | [-16,1 -> var.] | |
| WS7 | Source | Dec.1 | 297 | -15,9 | Dec.9 | WS7 | Dec.1-9 | -25 | -14,8 | 0,8 |
| | | | | | | WS7 | Dec.12-26 | -37 | -14,1 | 0,66 |
| WS8 | I/med. | Dec.7 | 286 | -14,2 | Dec.11 | (n.d.) | | | | |
| WS9 | Front | Dec.8 | 272 | -14,7 | Dec.22 | WS9 | Dec.8-11 | [~-10] | -14,1 | 0,67 |
| | | | | | | WS9 | Dec.12-18 | -66 | -11,7 | 0,45 |
| WS9a | Front | Dec.22 | 254 | -15,0 | Dec.26+ | WS9a | Dec.20-24 | -42 | -15,1 | 0,26 |
| WS10 | Source | Dec.10 | 299 | -16,1 | Jan.5+ | WS10 | Dec.10-14 | 0 | -16,1 | 0,26 |
| | | | | | | WS10 | Dec.14-Jan.5 | -41,5 | [-14,9 -> -13,5] | |
| WS11 | I/med. | Dec.13 | 287,5 | -14,8 | Jan.5+ | (n.d.) | | | | |
| WS12 | Source | Dec.20 | 302 | nd | Jan.5+ | WS12 | Dec.23-Jan.5 | -45 | -14,3 | 0,63 |
| WS13 | Source | Jan.3 | 293 | nd | Jan.23+ | (n.d.) | | | | |
| WS9b* | Front | Jan.3 | 240 | nd | Jan.6+ | WS9b | Dec.30-Jan.4 | -26 | -11,2 | 0,39 |

*Columns are as follows:*
Number of spot; Location; Date of appearance; Mean L2 and latitude (zenographic) from first two rotations;
Date last seen as a distinct white spot; Dates tracked (sometimes including the light cell after the plume had faded);
Mean DL2 (degrees per 30 days) and latitude; Standard deviation for latitude (some are unusually high due to irregular shape and migration).
*WS9b was not numbered in interim reports. It arose within an existing (not methane-bright) white streak at the front, which may have been 9a.

**Table 2:  Latitudes and speeds of retrograding dark spots on SEBs.**

| | L2 | Dates | DL2 | Lat. | SD | n |
|---|---|---|---|---|---|---|
| **Dark spots in S. branch:** | | | | | | |
| | P1-P3 | Nov.16-25 | 132 | -20,7 | 0,60 | 41 |
| | P1-P3 | Nov.27-Dec.4 | 117 | -20,8 | 0,40 | 11 |
| | P1 | Dec.3-23 | 110 | -20,8 | 0,32 | 7 |
| | DS1 | Nov.24-Dec.6 | 118 | -21,2 | 0,24 | 13 |
| | DS3 | Nov.28-Dec.6 | 120 | -20,0 | 0,45 | 9 |
| | DS3 | Dec.3-31 | 114 | -20,2 | 0,50 | 4 |
| | DS4 | Dec.3-7 | 114 | -20,1 | 0,30 | 3 |
| | DS5 | Nov.27-Dec.6 | 108 | -20,6 | 0,40 | 9 |
| | DS5 | Dec.7-26 | 45 | -21,9 | 0,55 | 4 |
| | DS6 | Dec.5-19 | 116 | -20,0 | 0,59 | 3 |
| | A (DS3) | Dec.31-Jan.18 | 108 | -21,3 | 0,20 | 3 |
| | B (DS1)* | Dec.19-Jan.16 | 57 | -22,1 | 0,69 | 4 |
| | B (DS1) | Dec.31-Jan.6 | 80 | -21,8 | 0,70 | 2 |
| | B (DS1) | Jan.8-16 | 35 | -22,4 | 0,30 | 2 |
| | *B: Remnant of DS1 after passing DS3.  DL2 oscillating between +35 and +80. | | | | | |
| | *[Un-named northerly* | Dec.4-13 | 40 | -19,6 | 0,30 | 11 |
| | *spots between ovals]* | Dec.31-Jan.8 | 51 | -20,0 | 0,33 | 8 |
| | E,F,G (merging) | Dec.31-Jan.12 | 44 | -21,8 | 0,54 | 13 |
| | F | Jan.11-17 | ~0 | -22,2 | 0,43 | 7 |
| | K | Jan.4-8 | 134 | -21,0 | | 1 |
| | M | Jan.2-17 | 70 | -21,9 | 0,47 | 9 |
| | P | Jan.2-9 | 72 | -21,1 | 0,17 | 3 |
| | P | Jan.10-Feb.8 | 34 | -22,3 | 0,83* | 6 |
| | T | Jan.4-12 | 56 | -21,9 | 0,50 | 4 |
| | T (v.large) | Jan.13-Feb.7 | ~+44 to -16 *(sic)* | | | |
| | | | | -22,8 | (Jan.18) | 1 |
| | J1 | Jan.23-Feb.13 | 43 | -21,7 | 0,74* | 7 |
| | J4 | Feb.6-13 | 43 | -21,4 | 0,77* | 7 |
| | *Latitude measurements for these slowly-retrograding spots/streaks were bimodal, | | | | | |
| | possibly measured as: (Streak on S edge:) | | | -22,2 | 0,35 | 8 |
| | | (Streak + dark belt:) | | -20,8 | 0,23 | 6 |
| **Chains of white ovals in SEB(S):** | | | **DL2** | Lat. | SD | n |
| Sector 0 | (7 dark projs. | Dec.11-Jan.2 | 69,9 | | | |
| | between 8 ovals) | | (+/-4,3) | | | |
| | [Ahead of DS3/A.] | | | | | |
| Sector 1 | (7 ovals tracked | Jan.11-Feb.1 | 50,6 | -20,5 | 0,36 | 17 |
| | out of 8 present) | | (+/-4,5) | | | |
| | [Ahead of d.s.F.] | | | | | |
| Sector 2 | (3 ovals tracked | Jan.9-Feb.8 | 47,7 | -20,3 | 0,39 | 11 |
| | out of 5-7 present) | | (+/-6,4) | | | |
| | [Alongside the major dark spots..] | | | | | |

The numerical columns are as follows:  DL2 (degrees per 30 days); Latitude; Standard deviation for latitude;
n, number of latitude measurements (small in some cases due to difficulty isolating the spots).
As shown in the chart, drift rates were determined from a larger number of longitude measurements (done manually).

# Table 3: Summary of key features of SEB Revival

| | | | DL2 (deg/mth) | | u₃ (m/s) | Lat. (zenographic) | | No.of spots |
|---|---|---|---|---|---|---|---|---|
| | | *Dates* | *Mean* | *SD or [Range]* | *Mean* | *Mean* | *SD or [Range]* | *tracked* |
| **Start** (initial w.s.) | | Nov.9 | -- | | -- | -17,1 | | 1 |
| **Central branch:** | | | | | | | | |
| Source (locus of new plumes) | WS1-3: | Nov.9-17 | -- | | -- | -17,3 | 0,2 | 3 |
| | WS3-12: | Nov.17-Dec.20 | 5,3 | | -6,2 | -16,1 | 0,2 | 4 |
| White spots (plumes, | Hi-lat: | Nov.9-Jan.5 | -3,0 | 6,2 | -2,3 | -16,9 | 0,7 | 7 |
| arising at source or leading edge) | Lo-lat: | Nov.12-Jan.5 | -33,3 | 15,9 | 11,8 | -14,0 | 1,1 | 8 |
| Leading edge | | Nov.21-Feb.18 | -32 | [~-42 to 0] | 11,2 | (-14) | | |
| **S. branch:** | | | | | | | | |
| Pre-existing chain | D.ss: | Nov.1-Jan.2 | 72 | [+66 to +78] | -36,3 | -20,5 | 0,5 | many |
| | W.ss: | (ditto) | | | | -20,4 | 0,4 | many |
| Fast dark spots in S.branch: | | Nov.16-Jan.18 | 117,4 | 8,7 | -56,9 | -20,6 | 0,5 | 11 |
| | | | | [+108 to 134] | | | [-20.0 to -21.3] | |
| Slower dark spots in S.branch | | Dec.7-Feb.13 | 49,9 | 11,9 | -26,0 | -22,0 | 0,2 | 10 |
| (on S edge of SEB(S), omitting extremes) | | | | [~0 to +70] | | | [-21.9 to-22.3] | |
| Renewed chain of w.ovals | | Jan.9-Feb.8 | 49,7 | 5 | -26,2 | -20,4 | 0,4 | 10 |
| in revived SEB(S) | | | | [+43 to +57] | | | | |
| **N.branch:** | | | | | | | | |
| First major d.s. | | Jan.4-Feb.3 | -130 | | 57,7 | -11,0 | 0,5 | 1 |
| All dark spots | | Jan.5-Feb.9 | -120 | [~-150 to -90] | 53,0 | -10,7 | 0,9 | 12 |

# Tables and Figures (reduced copies)

Table 1.  Life histories of the bright plumes and cells.

| Table 1 | Life histories of the bright plumes and cells | | | | | | | | | |
|---|---|---|---|---|---|---|---|---|---|---|
| *No.* | *Location* | *First observations:* | | | *Last (w.s.):* | | *Track (cell):* | | | |
| | | *Date* | *L2* | *Lat.* | *Date* | | *Dates* | *DL2* | *Lat.* | *SD* |
| WS1 | Source | Nov.9 | 288 | -17,1 | Nov.14 | WS1 | Nov.9-17 | -12 | -17,1 | 0,52 |
| WS-N | Front | Nov.12 | 286 | -14,0 | (-->WS5 ) | WS-N | Nov.12-27 | -17 | -13,6 | 0,69 |
| WS2 | Source | Nov.14 | 291 | -17,5 | Nov.20 | WS2 | Nov.12-21 | -6 | [-17,5 -> -14] | |
| WS3 | Source | Nov.17 | 294,5 | -17,3 | Dec.13 | WS3 | Nov.17-25 | 4 | [-17,2 -> -15,4] | |
| | | | | | | WS3 | Nov.26-31 | -53 | -14,3 | 0,46 |
| | | | | | | WS3 | Dec.2-14 | -11 | -12,1 | 0,53 |
| WS4 | Front | Nov.19 | 290 | -19,4 | Nov.24 | WS4 | Nov.20-29 | -3 | -18,9 | 0,55 |
| WS5 | Front | Nov.23 | 284 | -16,6 | Dec.12 | WS5 | Nov.24-Dec.12 | -6 | -17,4 | 0,62 |
| WS6 | Front | Nov.28 | 279 | -16,1 | Dec.8 | WS6 | Nov.28-Dec.8 | -6 | [-16,1 -> var.] | |
| WS7 | Source | Dec.1 | 297 | -15,9 | Dec.9 | WS7 | Dec.1-9 | -25 | -14,8 | 0,8 |
| | | | | | | WS7 | Dec.12-26 | -37 | -14,1 | 0,66 |
| WS8 | I/med. | Dec.7 | 286 | -14,2 | Dec.11 | (n.d.) | | | | |
| WS9 | Front | Dec.8 | 272 | -14,7 | Dec.22 | WS9 | Dec.8-11 | [~-10] | -14,1 | 0,67 |
| | | | | | | WS9 | Dec.12-18 | -66 | -11,7 | 0,45 |
| WS9a | Front | Dec.22 | 254 | -15,0 | Dec.26+ | WS9a | Dec.20-24 | -42 | -15,1 | 0,26 |
| WS10 | Source | Dec.10 | 299 | -16,1 | Jan.5+ | WS10 | Dec.10-14 | 0 | -16,1 | 0,26 |
| | | | | | | WS10 | Dec.14-Jan.5 | -41,5 | [-14,9 -> -13,5] | |
| WS11 | I/med. | Dec.13 | 287,5 | -14,8 | Jan.5+ | (n.d.) | | | | |
| WS12 | Source | Dec.20 | 302 | nd | Jan.5+ | WS12 | Dec.23-Jan.5 | -45 | -14,3 | 0,63 |
| WS13 | Source | Jan.3 | 293 | nd | Jan.23+ | (n.d.) | | | | |
| WS9b* | Front | Jan.3 | 240 | nd | Jan.6+ | WS9b | Dec.30-Jan.4 | -26 | -11,2 | 0,39 |

*Columns are as follows:*

Number of spot; Location; Date of appearance; Mean L2 and latitude (zenographic) from first two rotations;

Date last seen as a distinct white spot; Dates tracked (sometimes including the light cell after the plume had faded);

Mean DL2 (degrees per 30 days) and latitude; Standard deviation for latitude (some are unusually high due to irregular shape and migration).

*WS9b was not numbered in interim reports. It arose within an existing (not methane-bright) white streak at the front, which may have been 9a.

## Table 2: Latitudes and speeds of retrograding dark spots on SEBs.

| | L2 | Dates | DL2 | Lat. | SD | n |
|---|---|---|---|---|---|---|
| **Dark spots in S. branch:** | | | | | | |
| | P1-P3 | Nov.16-25 | 132 | -20,7 | 0,60 | 41 |
| | P1-P3 | Nov.27-Dec.4 | 117 | -20,8 | 0,40 | 11 |
| | P1 | Dec.3-23 | 110 | -20,8 | 0,32 | 7 |
| | DS1 | Nov.24-Dec.6 | 118 | -21,2 | 0,24 | 13 |
| | DS3 | Nov.28-Dec.6 | 120 | -20,0 | 0,45 | 9 |
| | DS3 | Dec.3-31 | 114 | -20,2 | 0,50 | 4 |
| | DS4 | Dec.3-7 | 114 | -20,1 | 0,30 | 3 |
| | DS5 | Nov.27-Dec.6 | 108 | -20,6 | 0,40 | 9 |
| | DS5 | Dec.7-26 | 45 | -21,9 | 0,55 | 4 |
| | DS6 | Dec.5-19 | 116 | -20,0 | 0,59 | 3 |
| | A (DS3) | Dec.31-Jan.18 | 108 | -21,3 | 0,20 | 3 |
| | B (DS1)* | Dec.19-Jan.16 | 57 | -22,1 | 0,69 | 4 |
| | B (DS1) | Dec.31-Jan.6 | 80 | -21,8 | 0,70 | 2 |
| | B (DS1) | Jan.8-16 | 35 | -22,4 | 0,30 | 2 |
| | *B: Remnant of DS1 after passing DS3. DL2 oscillating between +35 and +80. | | | | | |
| | [Un-named northerly | Dec.4-13 | 40 | -19,6 | 0,30 | 11 |
| | spots between ovals] | Dec.31-Jan.8 | 51 | -20,0 | 0,33 | 8 |
| | E,F,G (merging) | Dec.31-Jan.12 | 44 | -21,8 | 0,54 | 13 |
| | F | Jan.11-17 | ~0 | -22,2 | 0,43 | 7 |
| | K | Jan.4-8 | 134 | -21,0 | | 1 |
| | M | Jan.2-17 | 70 | -21,9 | 0,47 | 9 |
| | P | Jan.2-9 | 72 | -21,1 | 0,17 | 3 |
| | P | Jan.10-Feb.8 | 34 | -22,3 | 0,83* | 6 |
| | T | Jan.4-12 | 56 | -21,9 | 0,50 | 4 |
| | T (v.large) | Jan.13-Feb.7 | ~+44 to -16 *(sic)* | | | |
| | | | | -22,8 | (Jan.18) | 1 |
| | J1 | Jan.23-Feb.13 | 43 | -21,7 | 0,74* | 7 |
| | J4 | Feb.6-13 | 43 | -21,4 | 0,77* | 7 |
| | *Latitude measurements for these slowly-retrograding spots/streaks were bimodal, | | | | | |
| | possibly measured as: (Streak on S edge:) | | | -22,2 | 0,35 | 8 |
| | (Streak + dark belt:) | | | -20,8 | 0,23 | 6 |
| **Chains of white ovals in SEB(S):** | | | **DL2** | **Lat.** | **SD** | **n** |
| Sector 0 | (7 dark projs. | Dec.11-Jan.2 | 69,9 | | | |
| | between 8 ovals) | | (+/-4,3) | | | |
| | [Ahead of DS3/A.] | | | | | |
| Sector 1 | (7 ovals tracked | Jan.11-Feb.1 | 50,6 | -20,5 | 0,36 | 17 |
| | out of 8 present) | | (+/-4,5) | | | |
| | [Ahead of d.s.F. ] | | | | | |
| Sector 2 | (3 ovals tracked | Jan.9-Feb.8 | 47,7 | -20,3 | 0,39 | 11 |
| | out of 5-7 present) | | (+/-6,4) | | | |
| | [Alongside the major dark spots..] | | | | | |

The numerical columns are as follows: DL2 (degrees per 30 days); Latitude; Standard deviation for latitude;
n, number of latitude measurements (small in some cases due to difficulty isolating the spots).
As shown in the chart, drift rates were determined from a larger number of longitude measurements (done manually).

# Table 3: Summary of key features of SEB Revival

| | | | DL2 (deg/mth) | | u₃ (m/s) | Lat. (zenographic) | | No. of spots |
|---|---|---|---|---|---|---|---|---|
| | | *Dates* | *Mean* | *SD or [Range]* | *Mean* | *Mean* | *SD or [Range]* | *tracked* |
| **Start** (initial w.s.) | | Nov.9 | -- | | -- | -17,1 | | 1 |
| **Central branch:** | | | | | | | | |
| Source (locus of new plumes) | WS1-3: | Nov.9-17 | -- | | -- | -17,3 | 0,2 | 3 |
| | WS3-12: | Nov.17-Dec.20 | 5,3 | | -6,2 | -16,1 | 0,2 | 4 |
| White spots (plumes, | Hi-lat: | Nov.9-Jan.5 | -3,0 | 6,2 | -2,3 | -16,9 | 0,7 | 7 |
| arising at source or leading edge) | Lo-lat: | Nov.12-Jan.5 | -33,3 | 15,9 | 11,8 | -14,0 | 1,1 | 8 |
| Leading edge | | Nov.21-Feb.18 | -32 | [~-42 to 0] | 11,2 | (-14) | | |
| **S. branch:** | | | | | | | | |
| Pre-existing chain | D.ss: | Nov.1-Jan.2 | 72 | [+66 to +78] | -36,3 | -20,5 | 0,5 | many |
| | W.ss: | (ditto) | | | | -20,4 | 0,4 | many |
| Fast dark spots in S.branch: | | Nov.16-Jan.18 | 117,4 | 8,7 | -56,9 | -20,6 | 0,5 | 11 |
| | | | | [+108 to 134] | | | [-20.0 to -21.3] | |
| Slower dark spots in S.branch | | Dec.7-Feb.13 | 49,9 | 11,9 | -26,0 | -22,0 | 0,2 | 10 |
| (on S edge of SEB(S), omitting extremes) | | | | [~0 to +70] | | | [-21.9 to-22.3] | |
| Renewed chain of w.ovals | | Jan.9-Feb.8 | 49,7 | 5 | -26,2 | -20,4 | 0,4 | 10 |
| in revived SEB(S) | | | | [+43 to +57] | | | | |
| **N.branch:** | | | | | | | | |
| First major d.s. | | Jan.4-Feb.3 | -130 | | 57,7 | -11,0 | 0,5 | 1 |
| All dark spots | | Jan.5-Feb.9 | -120 | [~-150 to -90] | 53,0 | -10,7 | 0,9 | 12 |

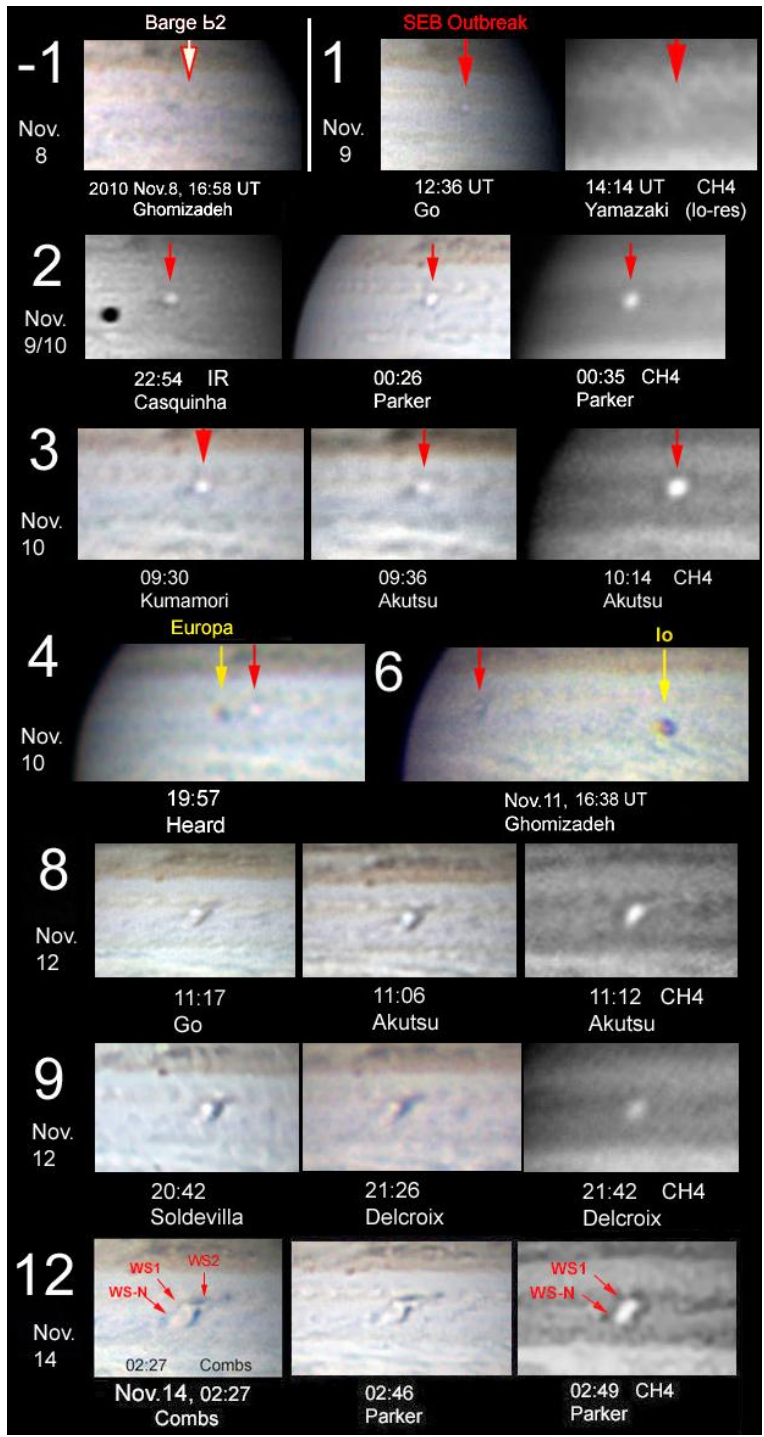

**Fig.1. Start of the outbreak.** The first image is from the previous day, showing barge b2, and subsequent images show the source during the first 12 rotations (numbered at left). Times are in UT, and observers' names are given. South is up in all figures; these panels cover approximately the latitudes from 30ºS to the equator. Images are in visible colour (RGB), unless otherwise stated, near-infrared (IR), and methane absorption band at 889 nm (CH4). The bright plume, WS1, is first detected as a tiny spot inside barge b2 on rotation 1. On rotation 6, the typical dark streak appears on its f. side. On rotation 7 (not shown), WS1 and the dark streak start expanding to the north. On rotation 8, a second white spot (WS-N) appears immediately N of WS1; it is initially barely detectable in methane band, but by rotation 12, it is as methane-bright as WS1. WS2 is first seen at the source on rotation 12. Europa and Io are seen in transit on rotations 4 and 6 respectively; their darkness has been exaggerated by the image processing.

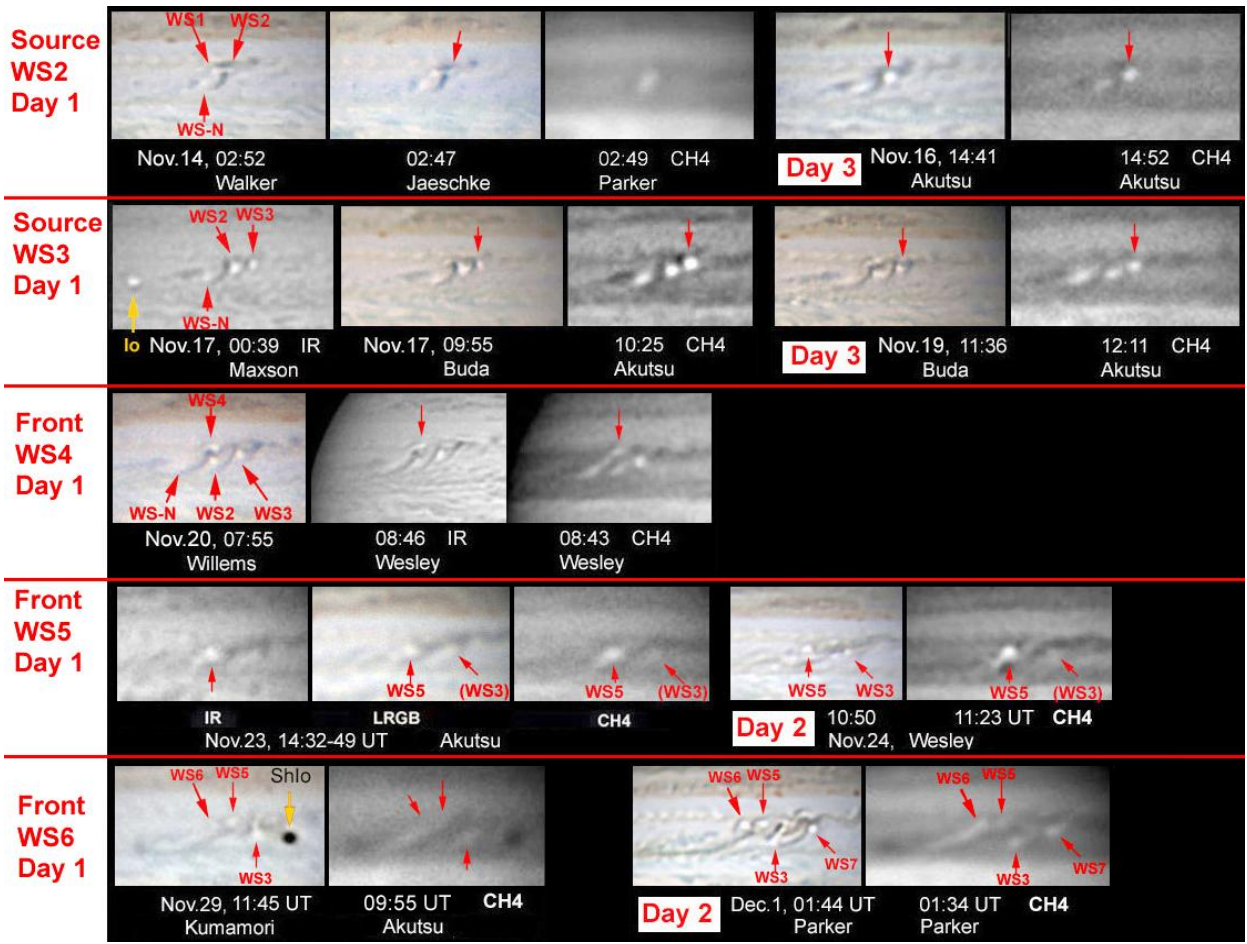

**Fig.2. Images of each of the bright plumes WS2 to WS6**, on its first and (when possible) third day, in visible light (RGB) and methane band (CH4). The last panel also shows WS7 at the source on its first day. (For plumes WS1 and WS-N, see **Fig.1**. A more complete gallery showing all plumes was posted in Report no.21.) Each plume was either at the f. edge (Source) or p. edge (Front) of the central branch. The images differ greatly in resolution and processing so the plume brightness cannot be compared quantitatively, especially in methane images. Also, apparent differences between plumes on the first day may reflect the very rapid growth that some showed even within 10-20 hours: WS1 and WS3 brightened especially quickly (see **Table 1**). Of the later plumes, not shown here, WS9a (at the front) and WS10 (at the source) were outstandingly methane-bright like the earlier ones, but others at the source (WS7 and WS12) were not.

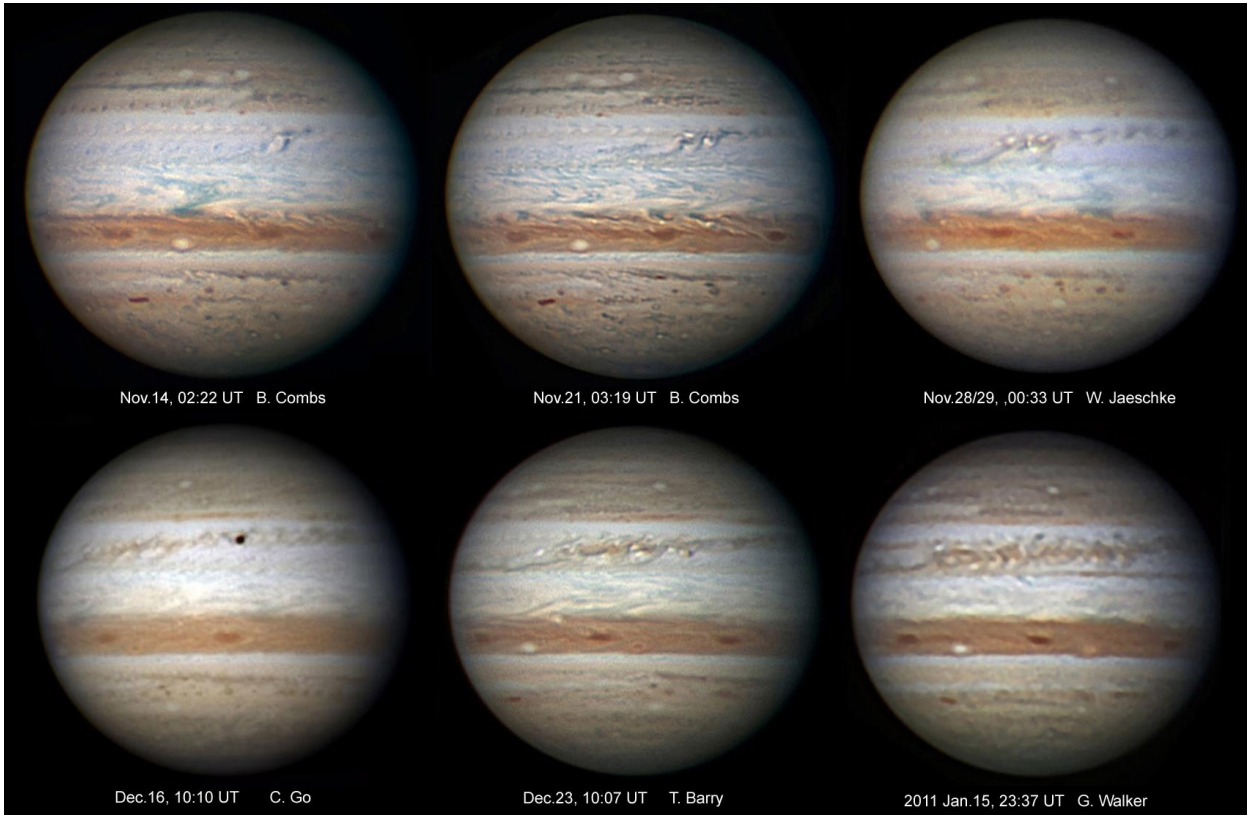

**Fig.3. Some of the best full-disk images showing the developing Revival.** Central meridian longitudes in System II are as follows. Nov.14: 269. Nov.21: 275. Nov.28/29: 297. Dec.16: 317. Dec.23: 286. Jan.15: 266.

*[NEXT PAGE]:*
**Fig.4. The overall structure of the Revival.**
 (A) The developing cellular structure of the central branch. Red arrows below indicate plumes at the source. Blue/cyan arrows above indicate plumes on the front (the leading edge complex). In the first image, the white arrowhead marks the cell created by WS7 which had disappeared but would soon reappear as WS11.
(B) Diagrammatic chart showing the evolution of the large cells derived from the bright plumes from the source (summarising Fig.5).
(C) Images and maps with the main components of the Revival labelled.
(D) Diagram of the typical structure of a developing SEB Revival [from ref.1].

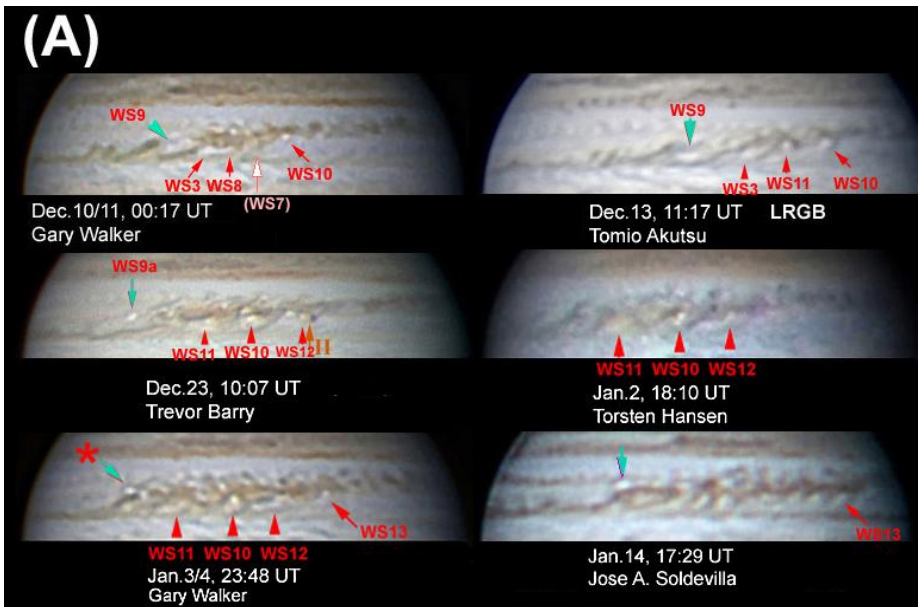
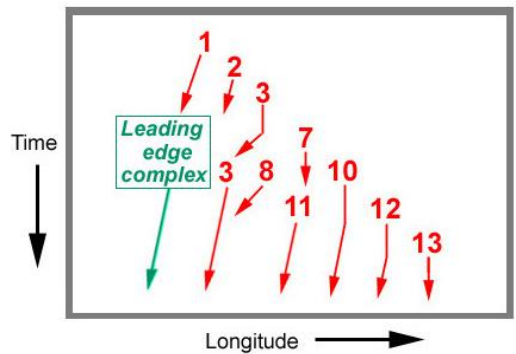
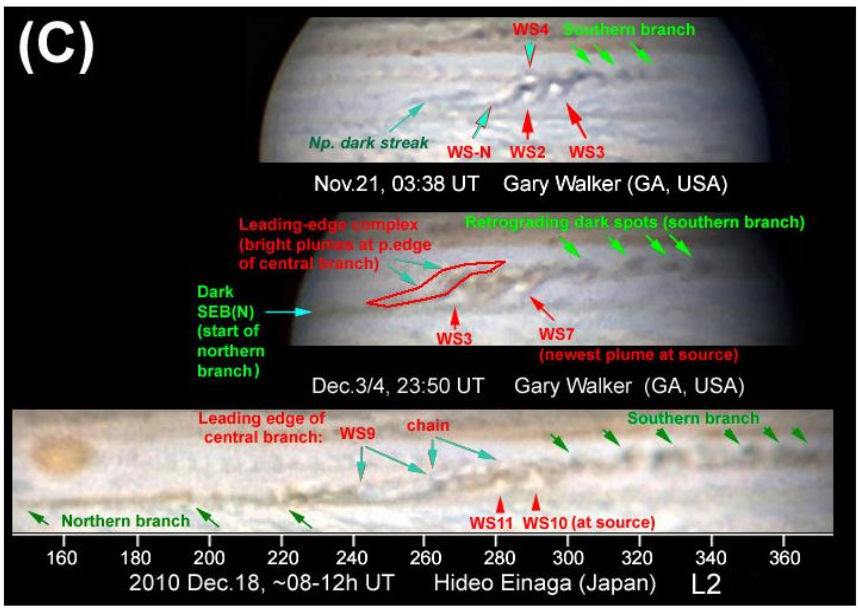
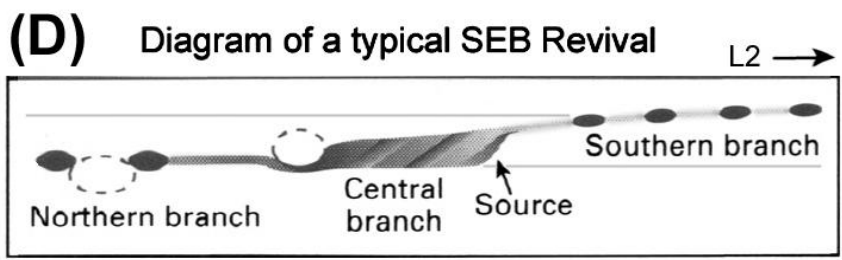

Fig.4

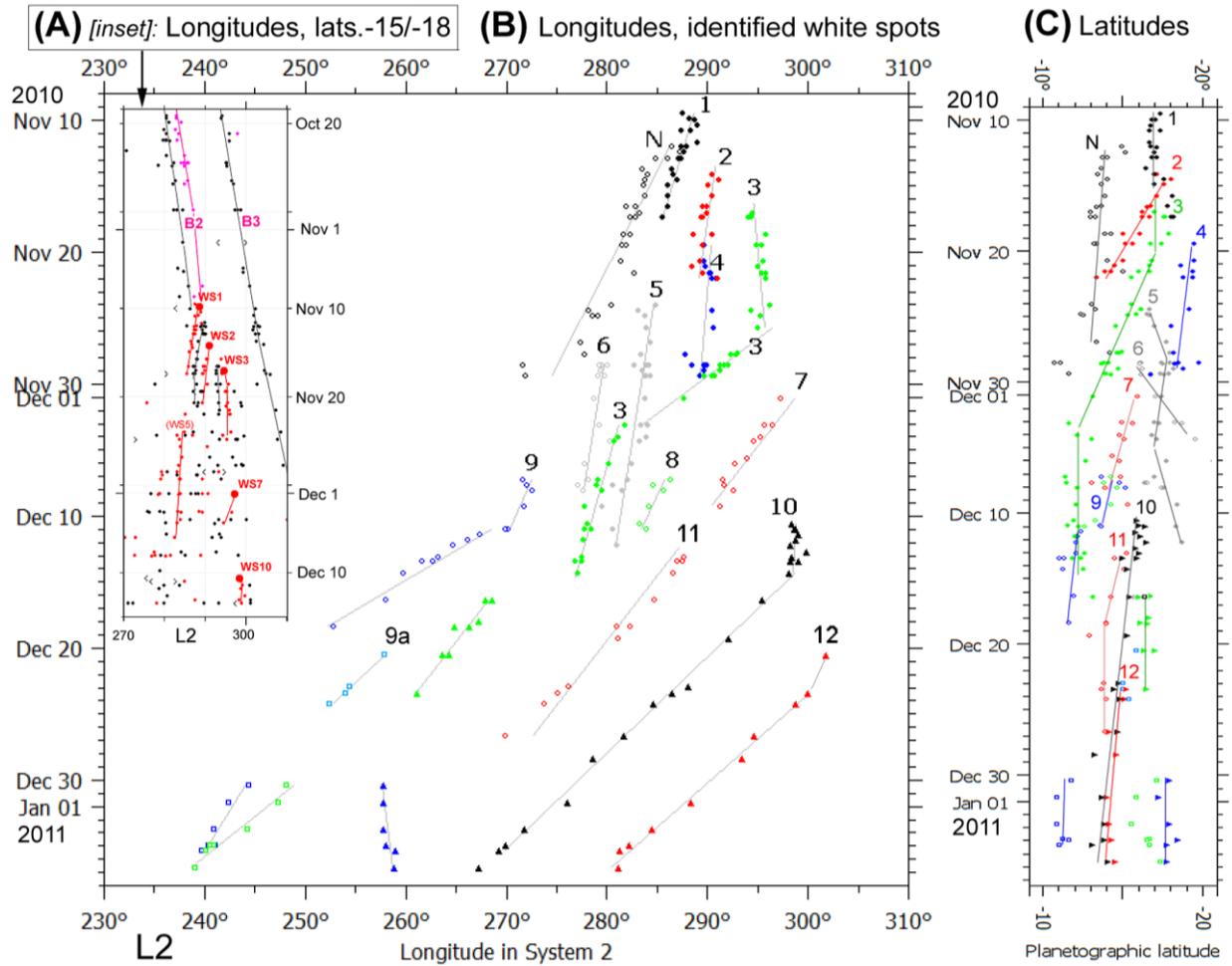

**Fig. 5. Charts of the bright plumes in the outbreak.**
**(A)** JUPOS chart of longitude vs time for bright and dark spots in the southern half of the SEB, latitude range 15 to 18ºS.  Pink, bright spot in barge b2; red, other bright spots, mainly plumes in the SEB Revival; black, dark spots (including the bluish p. rims of barges b2 and b3).  Large red dots are the first appearances of the plumes at the source, which all fall close to the extrapolated track of barge b2.
**(B)** Chart of longitude vs time for the bright spots (plumes) in the SEB Revival, produced from JUPOS data by Michel Jacquesson, colour-coded by spot number.
**(C)** Chart of latitude vs time for the same spots, produced by Michel Jacquesson.

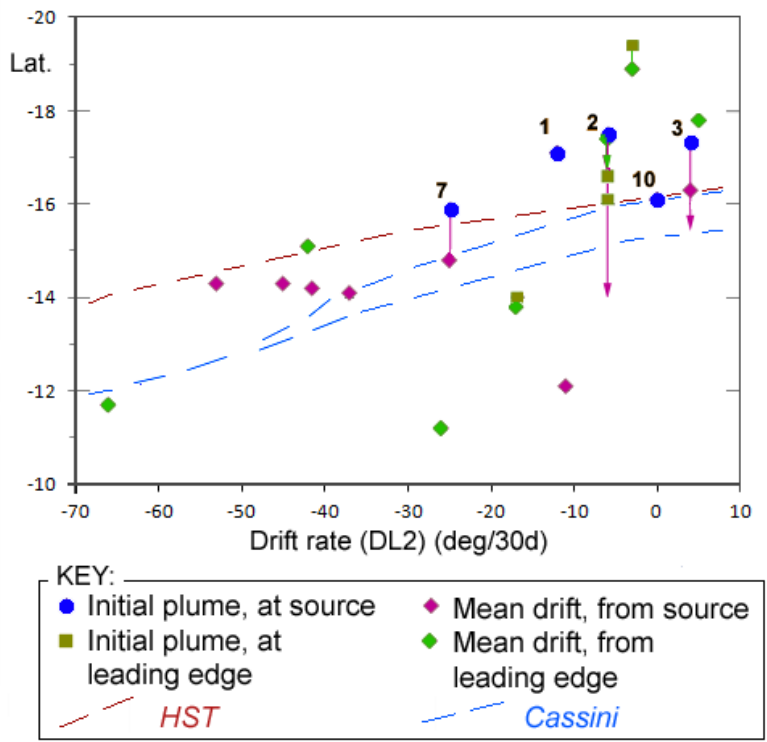

**Fig. 6. Chart of latitude vs speed for the bright spots** (plumes) in the SEB Revival, from data in **Fig.5 and Table 1**. Drift rates are typically ±1 to 3 deg/month, and latitudes typically ±0.5 to 0.6º (see **Fig.5 and Table 1**). Dashed lines are the mean zonal wind profiles from HST, 1994-1998 [ref.18] and from Cassini, 2000 [ref.30]. (Even in those 'normal' years, the ZWPs showed considerable scatter across this region, including multiple gradients as shown for Cassini, which may represent different longitude sectors). Positions of the plumes showed large real scatter, but notably, all above 16ºS and some below had drift rates close to System II. Magenta arrows indicate how some plumes migrated north without change of speed. Only spots below 14.5ºS developed rapid (negative) drifts in longitude.

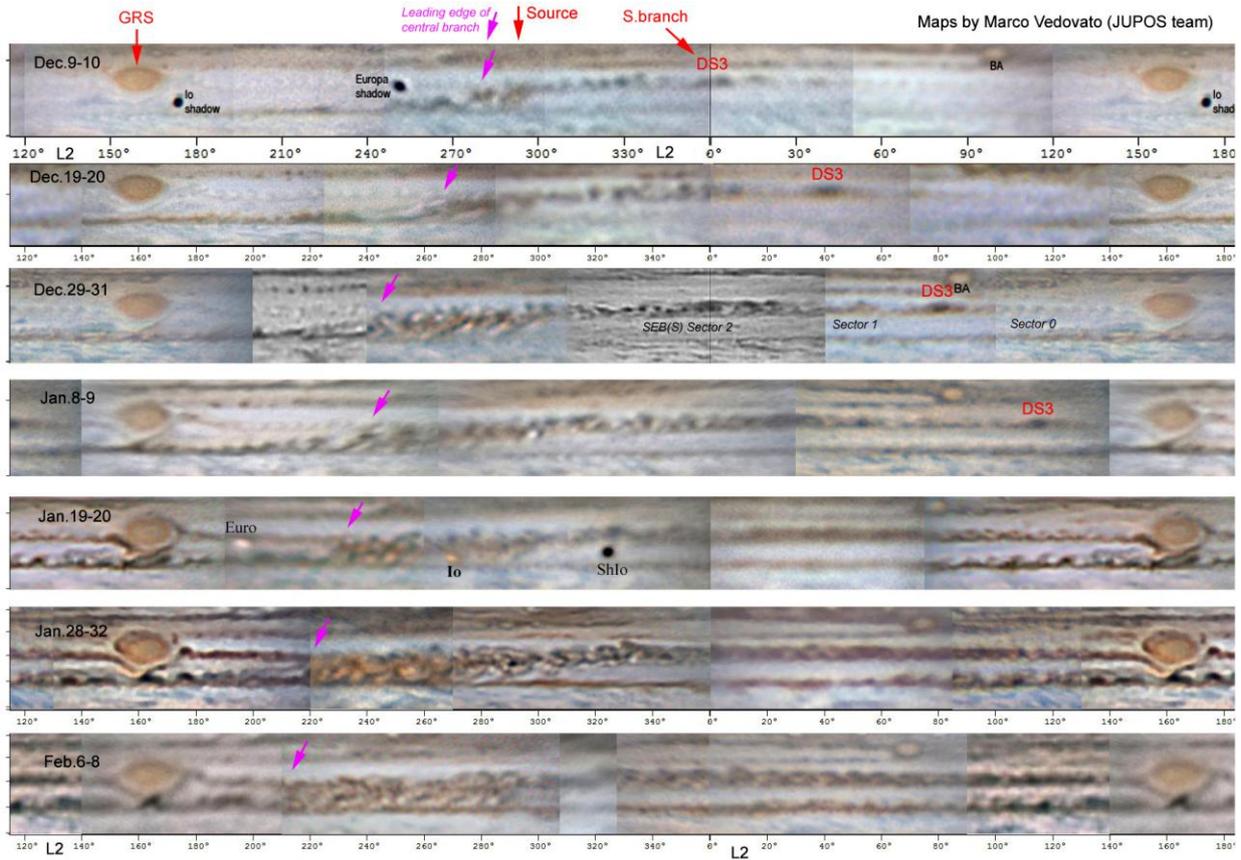

**Fig.7. Maps showing development of the SEB Revival, 2010 Dec.- 2011 Feb.** Indicated are the source (not moving much in L2), the leading edge of the central branch (magenta arrow), and the only dark spot in the southern branch that reached the GRS (DS3). Maps were made by Marco Vedovato using WinJUPOS, from images by the following observers (some of them from ALPO-Japan). Dec.9-10: Kumamori, Kazemoto, Combs, Willems. Dec.19-20: Chang, Barry, Kazemoto, Hatanaka. Dec.29-31: Willems, Combs, Wesley, Olivetti. Jan.8-9: Kidd, Akutsu. Jan.19-20: Put, Parker, Peach, Morales Rivera. Jan.28-Feb.1: Jolly, Parker, Tyler, G. Walker. Feb.6-8: Peach, Lasala, Akutsu, Parker.

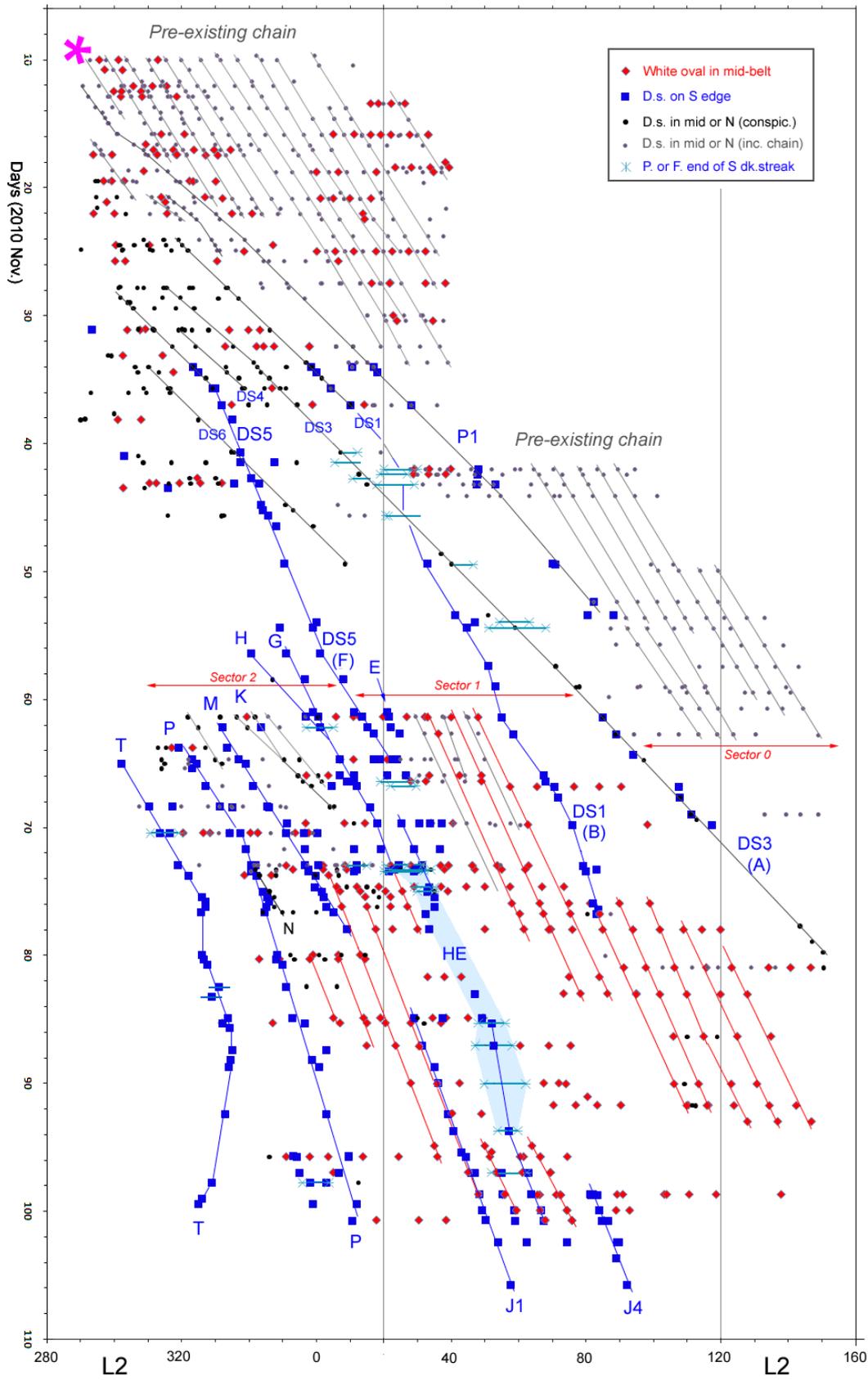

**Fig.8. Southern branch of the SEB Revival:** Complete chart of longitude (L2) vs time, for identified dark spots and for white ovals within SEB(S). Magenta asterisk indicates the initial outbreak.

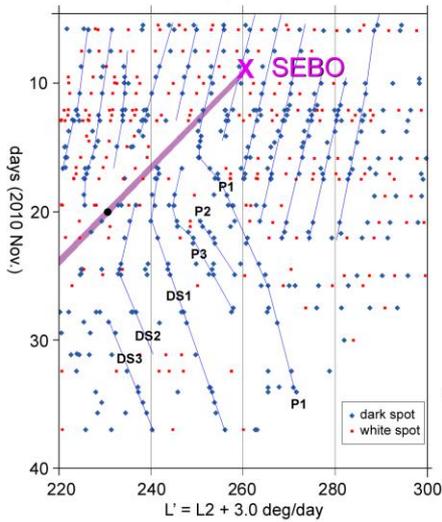
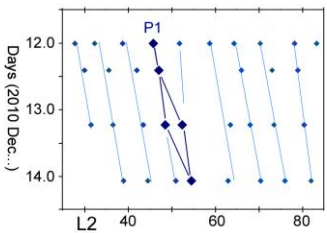
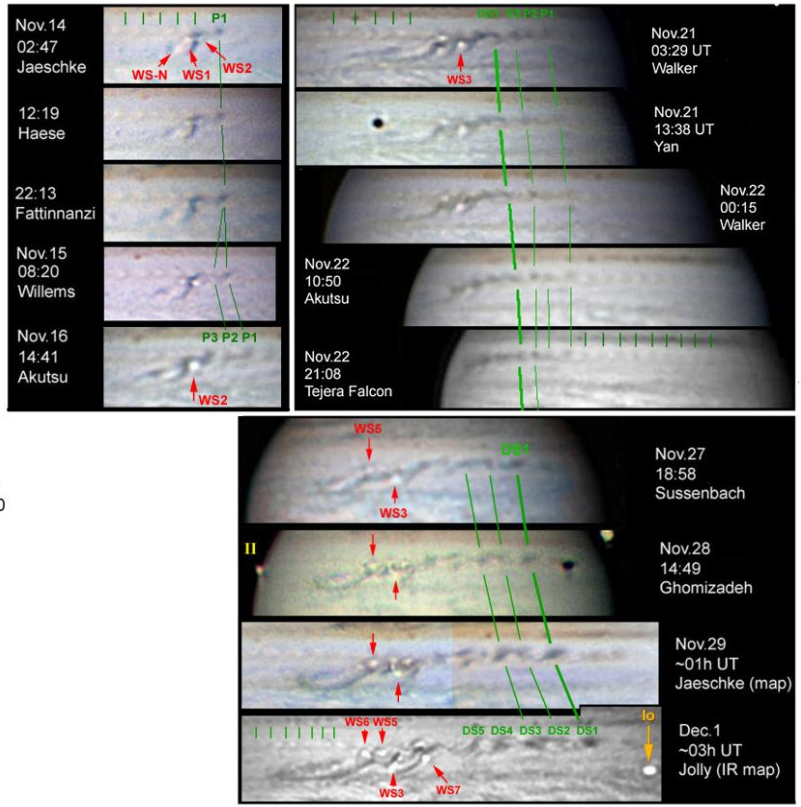

**Fig.9. Details of the dark spots in the southern branch.**
**(A,B)** Enlargements of the longitude-vs-time chart to show the relationship of the first rapidly retrograding dark spot (P1) to the pre-existing chain of grey projections. (A) shows how P1, and probably subsequent dark spots, arose by darkening and acceleration of the pre-existing projections. (A) has a longitude scale moving at +3.0 deg/day relative to System II, whereas other charts use System II longitude. (B) shows how P1 (large symbols) retained its appearance as a darker member of the pre-existing chain of small grey projections (small symbols) by shifting from one projection to the next within 2 days.
**(C)** Examples of images tracking these retrograding dark spots, Nov.14-Dec.1. These three panels show the development of dark spots P1 to P3 and DS1 to DS5. Rows of vertical green lines mark the pre-existing chain of grey projections in the first and/or last image of each panel. Some of the bright plumes are labelled in red.

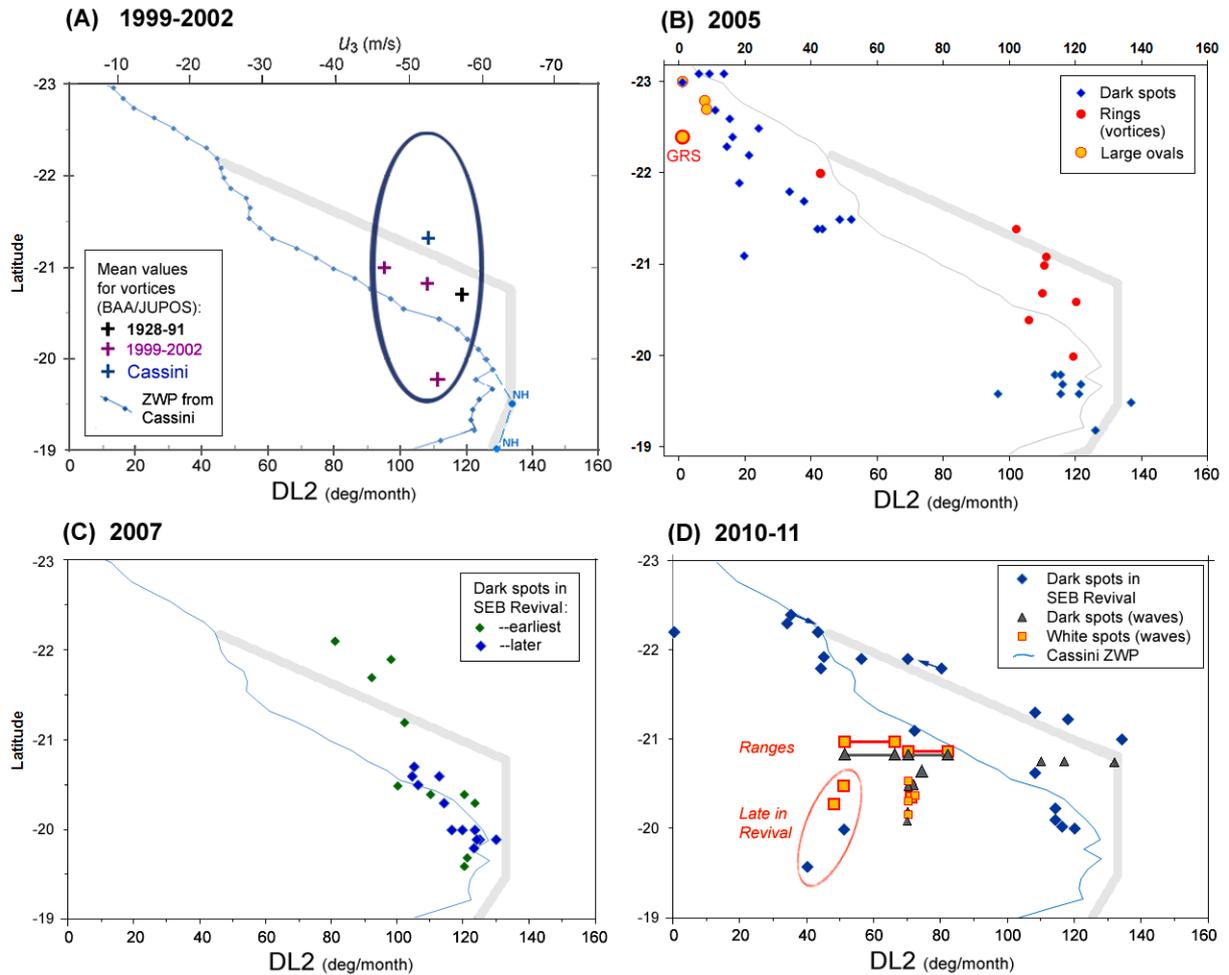

**Fig.10. Zonal drift profiles (ZDPs):** Charts of speed (DL2) vs latitude (zenographic) for SEB(S) spots, from BAA and JUPOS data, compared with the Cassini ZWP: adapted from Report no.22 [ref.8] and [ref.33]. Similar charts for 2010-2015 are published in [ref.20].

Data are plotted in DL2 (degrees per 30 days); a scale of wind speed $u_3$ (m/s in System III, as calculated for 20.8ºS) is given in (A). In each panel the blue line is the zonal wind profile (ZWP) from Cassini. The broad grey line is an estimated fit to the numerous points which have higher retrograding speeds than the ZWP for their latitude, viz. dark spots in the 2010 SEB Revival; vortices in all years shown here, and in 2011-2015 [see our on-line reports, & refs.20 & 33]; and two points from the New Horizons ZWP. This grey line could perhaps indicate a broader ZWP for the SEBs jet at deeper level.

**(A): Mean values for typical SEBs jetstream spots (vortices) during normal years of SEB activity**, from BAA, JUPOS, and Cassini data. Sources are as follows.
--1928-1991 (mean from Ref.1; includes SEB Revivals as well as normal years);
--1999-2002 (BAA reports in JBAA, from JUPOS data);
--2000 (Cassini: mean centre of vortices, estimated from publicly released maps, and their latitudinal extent indicated by the oval). Small blue points are the ZWP from Cassini [ref.30], plus two additional points from New Horizons [ref.31].

**(B) 2005:** from our final report [ref.34]

**(C) 2007:** dark spots in the SEB Revival: from our final report [ref.6]. Many of these spots had the circular forms and mutual interactions typical of vortices. Green points were the first such spots (re-analysed, confirming our original results with greater precision).

**(D) 2010:** southern branch of the SEB Revival: from Report no.24 [ref.8]. The chart icnludes the slow-moving pre-existing chain of grey and bright spots, and the dark spots in the southern branch, and slow-moving white ovals that reappeared within the SEB(S). Note that the chains of bright ovals and associated dark projections, both before and after the Revival, are systematically to the left of the ZWP from 19.5-21ºS, whereas the dark spots in the Revival are systematically to the right of the ZWP from 20.5-21.5ºS. Small arrows indicate spot B(DS1) which oscillated between alternate drift rates.

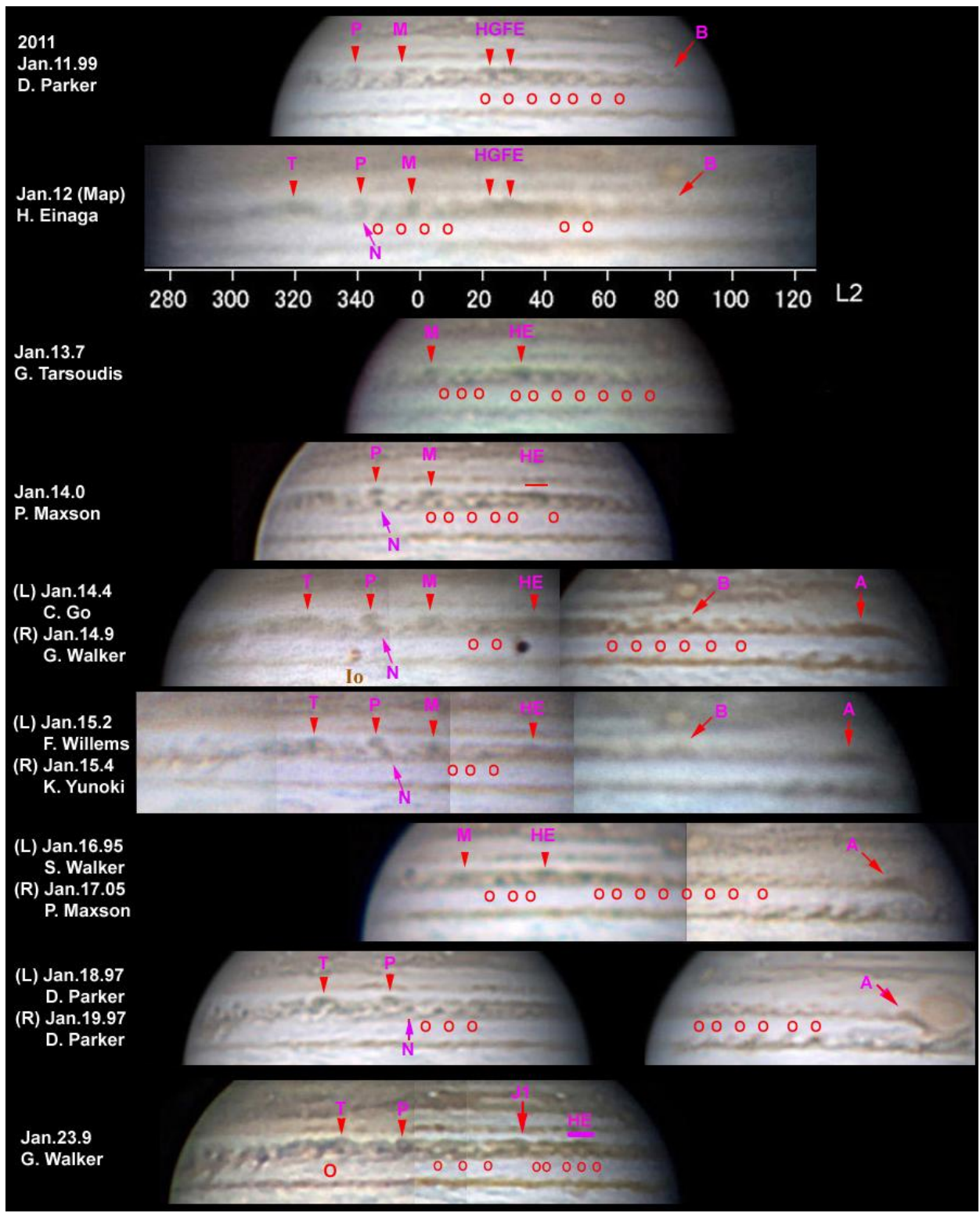

**Fig.11.** Images showing the southern branch, 2011 Jan.11-29, showing rapid and complex changes in the retrograding dark spots and streaks, and the recovering wave pattern on the SEBs jet (which is on the *north* edge of the reviving belt). Red arrowheads and arrows indicate the dark spots and streaks; red circles below the SEB(S) indicate the chain of small white ovals within the reviving belt, which manifested the recovering wave pattern. Images are approximately aligned in L2.

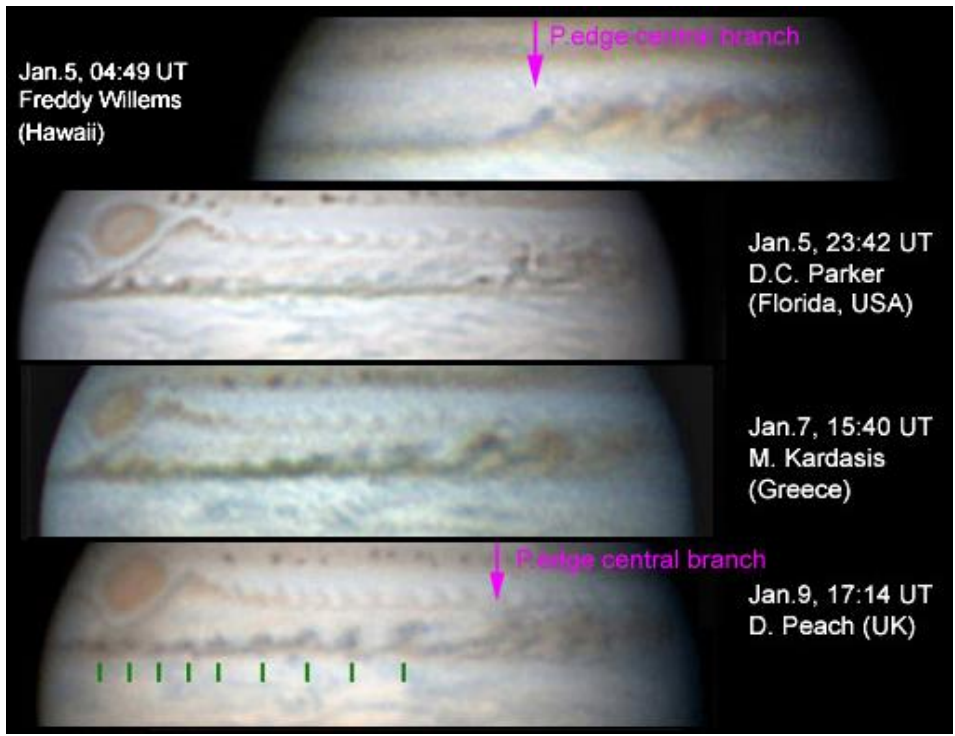
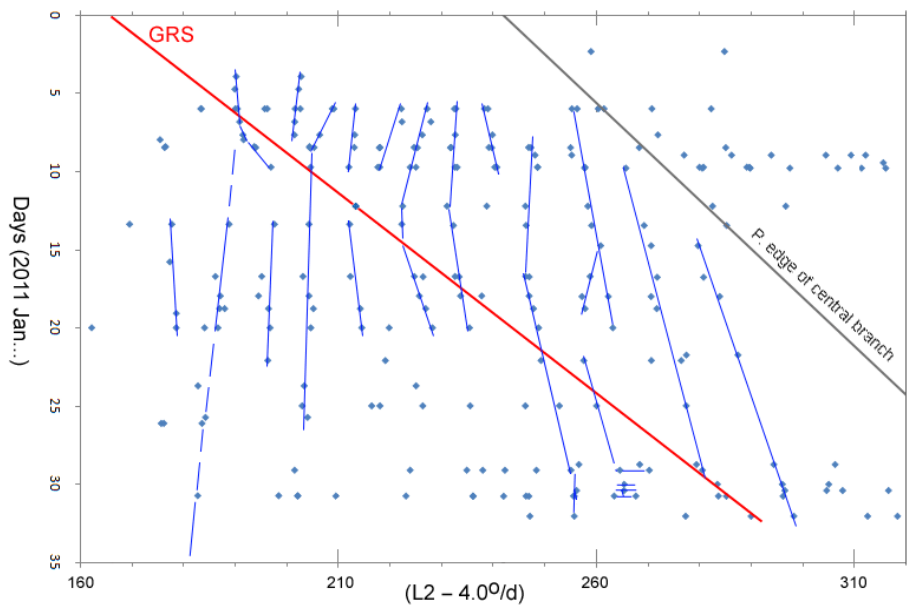

**Fig.12. Onset of the northern branch of the SEB Revival.**
**(A)** Images from 2011 Jan.5-9, showing the SEB(N) breaking up into very dark 'waves' (marked by dark green lines underneath) between the GRS and the leading edge of the central branch.
**(B)** Chart of longitude vs time. Longitude is in a system moving at -4.0 deg/day relative to System II.

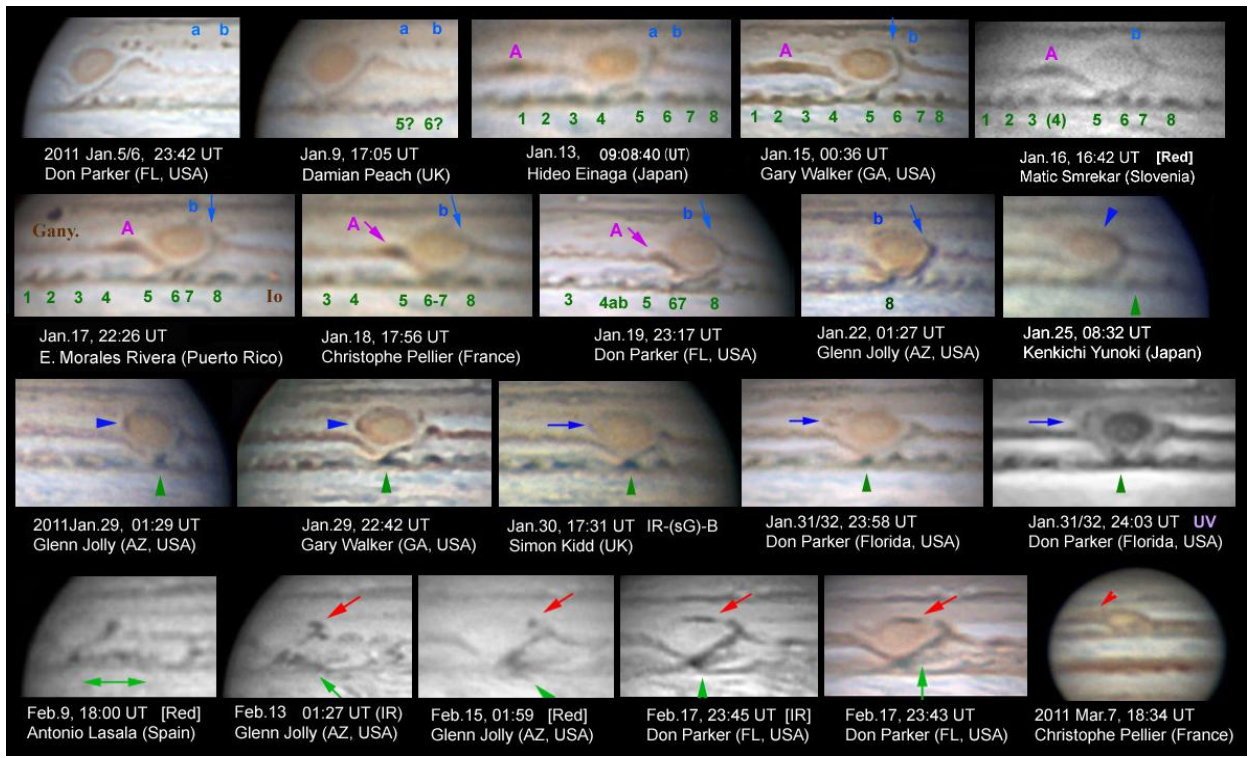

**Fig.13. Events at the GRS, 2011 January-March.**
**In January,** the dark spots on SEB(N) passed the GRS in succession, each becoming temporarily very dark as it did so. (Spots 1-8 are numbered in green; no.10 is indicated by a green arrowhead.) The RSH rim had already started to darken by Jan.5, and as the SEB(N) dark spots passed, dark material streamed rapidly anticlockwise around the GRS and also retrograded on the SEB(S) f. it. The bright plume N of the GRS was still visible in early Jan. but was displaced and was soon obliterated. Meanwhile, the faint leading edge of the S branch probably arrived around Jan.8, but the only distinct dark spot to arrive was A (=DS3), on Jan.18-19. This may have produced the especially dark streak which circumnavigated the GRS on Jan.22-29 and emerged p. it on Jan.30-31, but the SEB(N) and STBn spots may also have contributed. Two dark spots approached on the STBn jet: spot **a** slid N past the f. edge of the GRS (contributing to the developing dark collar: blue arrow), but **b** continued prograding past the S edge of the GRS and then probably disintegrated.
**In February** (red or IR images shown), very dark spots and streaks extended around the GRS, as the central branch of the Revival arrived from the f. side. On March 2, these appeared to be giving rise to the new S.Trop.Band. See text for description of indicated features.

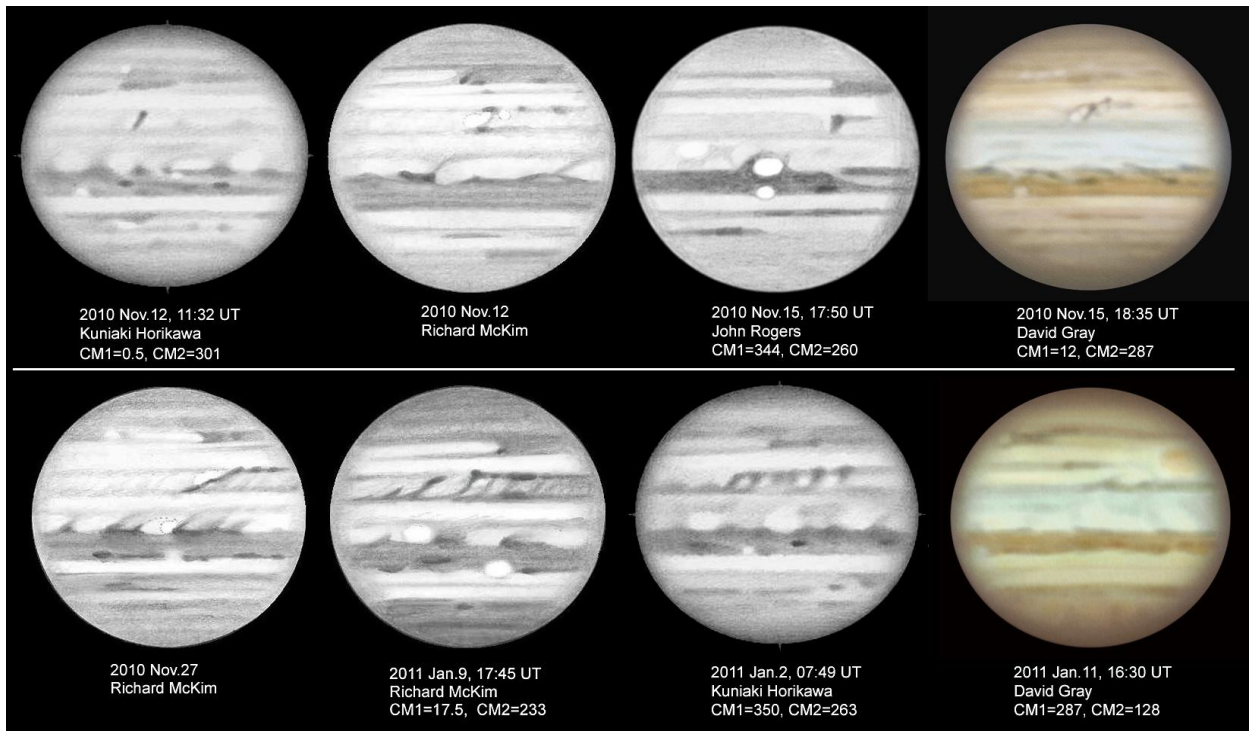

**Fig.14. Drawings of the SEB Revival.**
Visual observers were able to see features typical of historical Revivals, including the initial source with very dark 'column' and/or bright spot(s) (top row), the elaboration of the central branch (bottom row), and the southern branch advancing towards the GRS (bottom right).  Telescopes used were as follows: Kuniaki Horikawa (Japan), 30 cm reflector; Richard McKim (Northants., UK), 41-cm Dall-Kirkham; John Rogers (Cambs., UK), 25-cm reflector; David Gray (County Durham, UK), 41.5-cm Dall-Kirkham.